\documentclass[times,5p,twocolumn,final,nopreprintline]{elsarticle} 

\usepackage{framed,multirow}

\usepackage{amssymb}
\usepackage{latexsym}

\usepackage{url}
\usepackage{xcolor}

\usepackage[hidelinks]{hyperref}  

\usepackage{graphicx}
\usepackage{caption}
\usepackage{multirow}
\usepackage[flushleft]{threeparttable}
\usepackage{amsmath}
\usepackage{color, colortbl}

\newcolumntype{L}[1]{>{\raggedright\let\newline\\\arraybackslash\hspace{0pt}}m{#1}}
\newcolumntype{C}[1]{>{\centering\let\newline\\\arraybackslash\hspace{0pt}}m{#1}}
\newcolumntype{R}[1]{>{\raggedleft\let\newline\\\arraybackslash\hspace{0pt}}m{#1}}

\definecolor{newcolor}{rgb}{.8,.349,.1}

\begin{document}

\captionsetup[figure]{labelfont={bf},name={Fig.},labelsep=period}
\captionsetup[table]{labelfont={bf},labelformat={default},labelsep=period}


\begin{frontmatter}



\title{Superficial White Matter Analysis: An Efficient Point-cloud-based Deep Learning Framework with Supervised Contrastive Learning for Consistent Tractography Parcellation across Populations and dMRI Acquisitions}%




\author[1,2]{Tengfei~Xue}
\author[1]{Fan~Zhang\corref{cor1}}
\author[2]{Chaoyi~Zhang}
\author[1,2]{Yuqian~Chen}
\author[3]{Yang~Song}
\author[1]{Alexandra~J.~Golby}
\author[1,4]{Nikos~Makris}
\author[1]{Yogesh~Rathi}
\author[2]{Weidong~Cai}
\author[1]{Lauren~J.~O’Donnell\corref{cor1}}
\cortext[cor1]{Corresponding authors: fzhang@bwh.harvard.edu (F. Zhang) and odonnell@bwh.harvard.edu (L. J. O’Donnell).}

\address[1]{Brigham and Women’s Hospital, Harvard Medical School, Boston, USA}
\address[2]{School of Computer Science, University of Sydney, Sydney, Australia}
\address[3]{School of Computer Science and Engineering, University of New South Wales, Sydney, Australia}
\address[4]{Center for Morphometric Analysis, Massachusetts General Hospital, Boston, USA}

\begin{abstract}
Diffusion MRI tractography is an advanced imaging technique that enables \textit{in vivo} mapping of the brain’s white matter connections. White matter parcellation classifies tractography streamlines into clusters or anatomically meaningful tracts. It enables quantification and visualization of whole-brain tractography. Currently, most parcellation methods focus on the deep white matter (DWM), whereas fewer methods address the superficial white matter (SWM) due to its complexity. We propose a novel two-stage deep-learning-based framework, \textit{Superficial White Matter Analysis (SupWMA)}, that performs an efficient and consistent parcellation of 198 SWM clusters from whole-brain tractography. A point-cloud-based network is adapted to our SWM parcellation task, and supervised contrastive learning enables more discriminative representations between plausible streamlines and outliers for SWM. We train our model on a large-scale tractography dataset including streamline samples from labeled long- and medium-range (over 40~mm) SWM clusters and anatomically implausible streamline samples, and we perform testing on six independently acquired datasets of different ages and health conditions (including neonates and patients with space-occupying brain tumors). Compared to several state-of-the-art methods, SupWMA obtains highly consistent and accurate SWM parcellation results on all datasets, showing good generalization across the lifespan in health and disease. In addition, the computational speed of SupWMA is much faster than other methods. 
\end{abstract}


\end{frontmatter}


\section{Introduction}
\label{intro}

\begin{figure}[t] 
\centering
\includegraphics[width=0.46\textwidth]{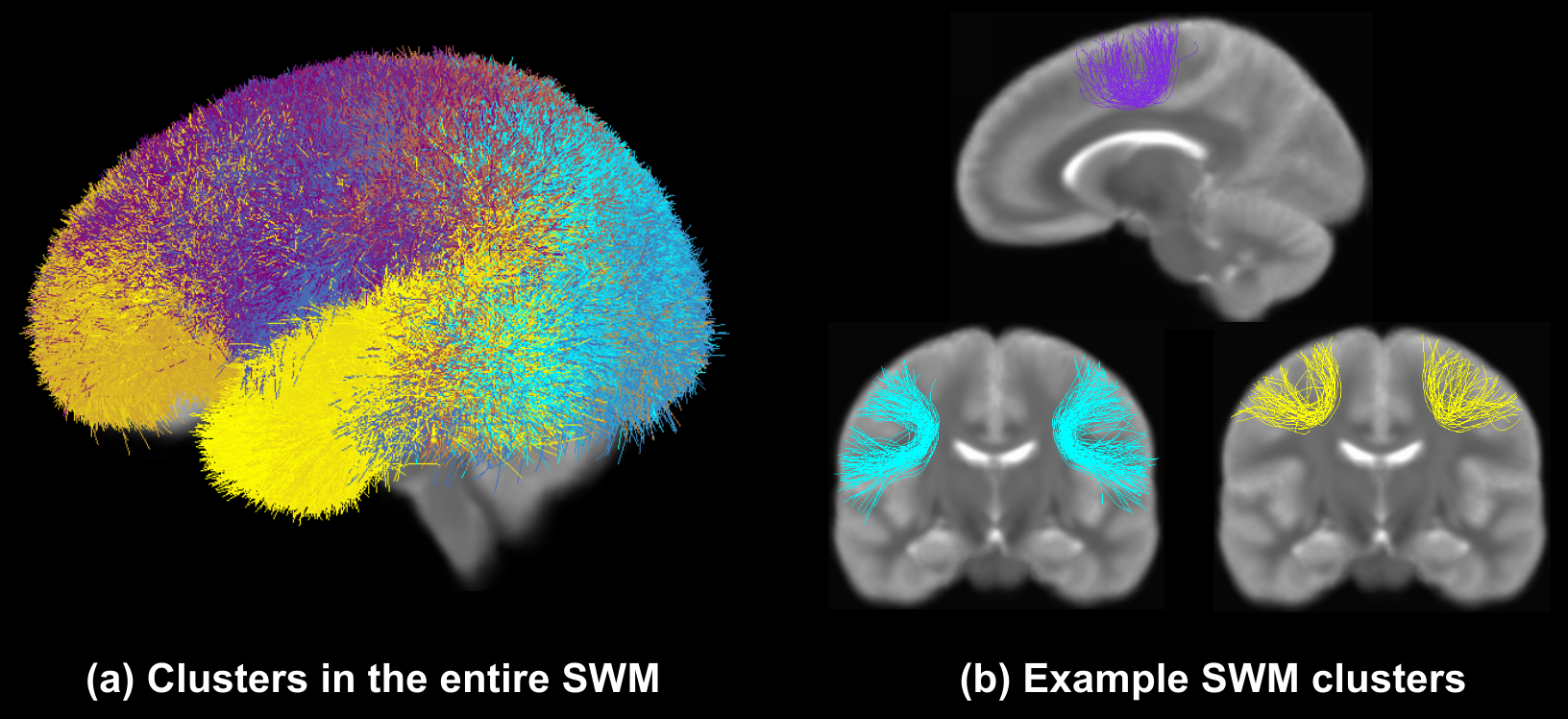}
\caption{\textmd{Visualization of SWM clusters. (a) shows clusters from the entire SWM, where each cluster has a unique color, and (b) shows several example SWM clusters.}}
\label{fig_swm_sample}
\end{figure}

Diffusion magnetic resonance imaging (dMRI) tractography is the only non-invasive method that can map the brain’s white matter connections \citep{Basser2000-tu}. Tractography methods estimate trajectories of brain white matter connections and represent those trajectories using streamlines. Each streamline is a set of ordered points in 3D space \citep{Zhang2022-sj}. Performing whole-brain tractography generates streamlines throughout the entire white matter, including the deep white matter (DWM) that connects distant cortical regions, and the superficial white matter (SWM) that includes short-range association connections (u-fibers) connecting adjacent and nearby gyri \citep{Guevara2020-da,Zhang2022-sj}. Whole-brain tractography can produce hundreds of thousands of streamlines, which are not directly useful to clinicians and researchers for quantification or visualization. Therefore, tractography parcellation is needed to classify white matter streamlines into clusters or anatomically meaningful tracts \citep{Zhang2022-sj}. Fig.~\ref{fig_swm_sample} provides a visualization of clusters in the entire SWM and several example SWM clusters from an anatomically curated atlas~\citep{Zhang2018-jx}, where clusters are annotated with anatomical labels by a neuroanatomist. 
SWM parcellation is important to enable analyses of the SWM in neuroscientific studies in health and disease \citep{Reginold2016-zq,DAlbis2018-ic,Malykhin2011-ex,Ji2019-ie} (see \citep{Guevara2020-da} for a review). 

Parcellation of the SWM is a challenging task due to the small diameter of SWM fasciculi, their high variability across brains, and their unique position near the cortex \citep{Guevara2020-da}. (We note that we focus on parcellation of SWM tractography in this work. Additional challenges inherent to performing SWM tractography, which include high curvature and complex fiber crossings \citep{Reveley2015-ki}, gyral bias \citep{Van_Essen2013-ka,Schilling2018-el}, and partial voluming effects \citep{Alexander2001-ai}, can be addressed to some extent by high-resolution imaging \citep{Song2014-un} and development of advanced tractography methods \citep{St-Onge2018-rw}.) Whereas most tractography parcellation methods currently focus on the DWM \citep{Yendiki2011-ea,Garyfallidis2018-hn,Wasserthal2018-if,Zhang2020-vm,Chen2021-dm}, few methods can parcellate the SWM \citep{Oishi2008-ct,Roman2017-uy,Guevara2020-da,Zhang2018-jx}. The existing SWM tractography parcellation methods use either region of interest (ROI)-based selection or streamline clustering. \textit{ROI-based methods} parcellate SWM tractography based on the ROIs streamlines end in and/or pass through \citep{Ouyang2016-kj,Oishi2008-ct,Malykhin2011-ex,Schilling2022-yz}. These ROI-based methods are the most commonly used but highly depend on the ROI parcellation scheme. \textit{Streamline clustering methods} group SWM streamlines based on the similarity of their geometric trajectories \citep{Zhang2018-jx,Guevara2012-ex,Roman2017-uy}. Streamline clustering methods for SWM parcellation are automatic and can leverage SWM atlases~\citep{Guevara2017-gz,Roman2017-uy,Zhang2018-jx,Roman2022-pr}, but challenges remain to achieve consistent parcellation across subjects and reduce runtime.

In recent years, deep-learning-based methods \citep{Wasserthal2018-if,Zhang2020-vm,Astolfi2020-zk,Xu2019-aa,Chen2021-dm,Chen2022-ow} have been successful for fast and consistent tractography parcellation, but representing dMRI data in a way that can best take advantage of deep networks is still an open challenge. \textit{Voxel-based} \citep{Wasserthal2018-if,Lu2021-hr,Wasserthal2019-ha} white matter tract parcellation methods take volumetric image data (e.g., fiber orientation distribution function (FOD) peaks \citep{Wasserthal2018-if}) and predict a tract's presence and/or orientation for each voxel. \textit{Streamline-based} tractography parcellation methods \citep{Zhang2020-vm,Xu2019-aa,Ngattai_Lam2018-ne} provide input to deep networks by encoding streamlines into different features. For example, DCNN+CL+ATT \citep{Xu2019-aa} uses a 1D feature descriptor containing streamline point spatial coordinates, and DeepWMA \citep{Zhang2020-vm} represents streamlines as 3-channel 2D images. However, the ambiguity of streamline data (the points along a streamline can equivalently be represented in forward or reverse order) poses challenges when using these representations as deep network input. \textit{Point clouds}, as an important geometric data format \citep{Guo2020-ql}, can potentially enable efficient and discriminative representations for streamlines. Coordinates of streamline points can be the input for point-cloud-based deep networks, as used in \citep{Astolfi2020-jf} for tractography filtering (binary classification of streamlines). There are two main advantages of using a point cloud representation for streamlines. First, streamline points are not on a regular grid and thus can be naturally represented using point clouds. Second, fiber tracking in tractography provides no directional information (points along a streamline can equivalently be represented in forward or reverse order). With the usage of a point cloud representation, streamlines with equivalent forward and reverse point orders (e.g., from cortex to subcortical structures or vice versa) can have the same global shape feature representation in the network. To our knowledge, no deep learning methods have focused on SWM parcellation, and point-cloud-based deep networks have not yet been used for white matter parcellation, in particular for SWM parcellation.

\begin{figure*}[t] 
\centering
\includegraphics[width=1.0\textwidth]{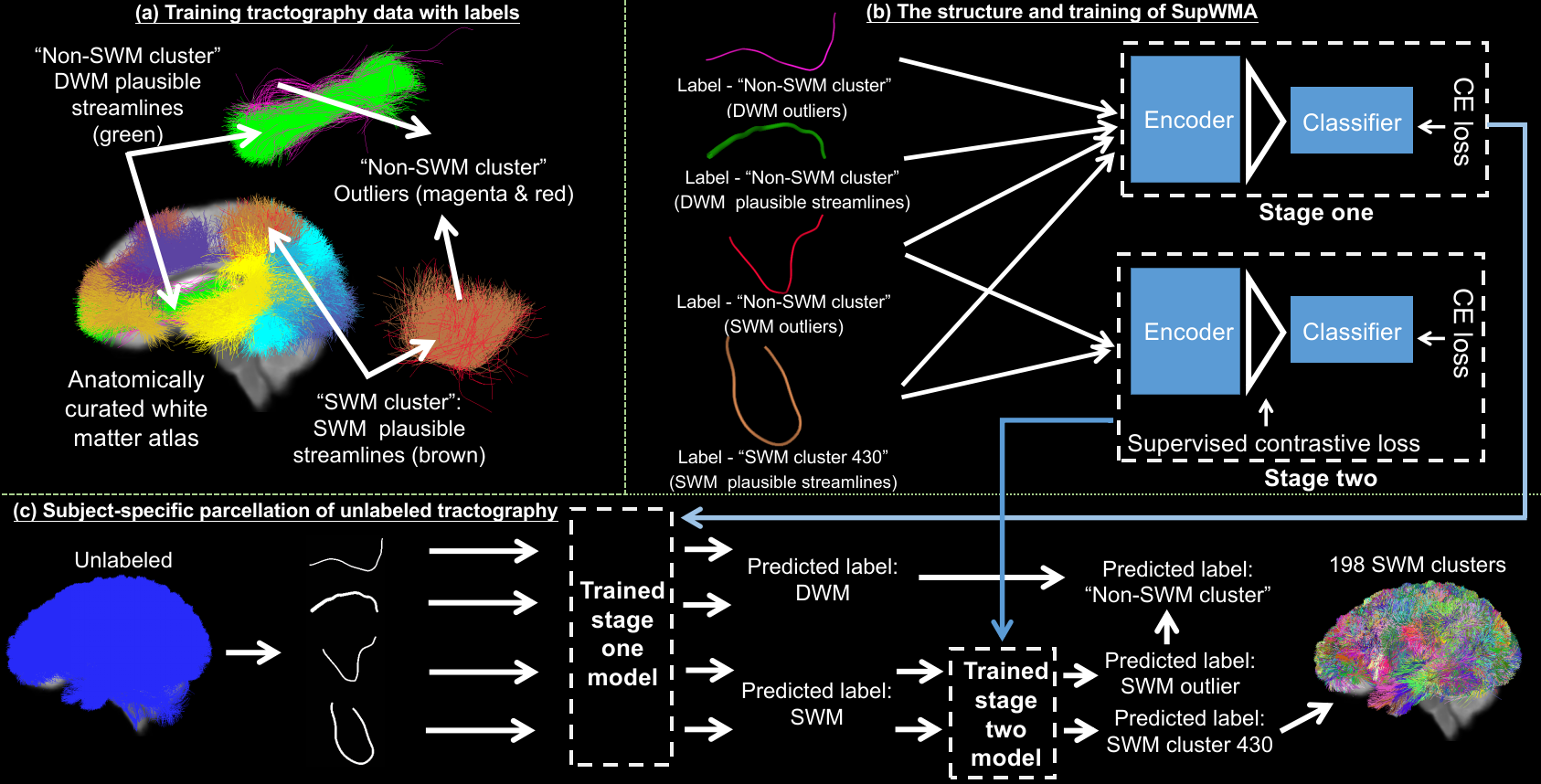}
\caption{\textmd{Overview of the SupWMA framework: (a) samples of training tractography data, (b) deep network structure and parcellation model training, (c) parcellation of unseen testing datasets.}}
\label{fig_overview}
\end{figure*}

In this paper, we propose a novel two-stage deep learning framework, 
\textit{Superficial White Matter Analysis (SupWMA)}, for SWM parcellation from whole-brain tractography. A point cloud classification network is used in both stage one and stage two. In stage one, the network takes streamlines of the whole-brain tractography as input, and outputs streamlines classified as SWM and DWM. In this way, stage one filters out DWM streamlines. In stage two, the network takes SWM streamlines as input, and it classifies them into 198 SWM clusters and outliers. Supervised contrastive learning is employed in stage two to obtain more discriminative representations of SWM streamlines. In our study, we focus on long- and medium-range SWM streamline connections (over 40~mm in length).

Our contributions are as follows. First, we decompose SWM parcellation into two stages, including a binary classification of SWM and DWM streamlines followed by a multi-class classification of 198 SWM clusters while performing removal of SWM outliers. Second, we modify the point-cloud-based network structure to preserve streamline pose and orientation information (because location in the brain is important for classification), and the modified network provides improvements to computational speed and accuracy. Third, we adapt supervised contrastive learning with a data augmentation technique (bilateral augmentation) for streamline classification. More discriminative representations between SWM plausible streamlines and outliers are obtained. Fourth, we evaluate our methods on six independently acquired datasets of 329 subjects with different ages and health conditions (including neonates and patients with space-occupying brain tumors), which are completely separate from the training set, and we obtain consistent results efficiently. 

This investigation extends our previous conference publication \citep{Xue2022-yz} to improve performance of SWM parcellation by designing a two-stage framework and incorporating bilateral data augmentation. Furthermore, we demonstrate successful SWM parcellation on three additional populations including neonates, patients with brain tumors, and patients with psychiatric disorders. We also include additional qualitative and quantitative evaluations of parcellation results, and an investigation of point importance along the streamline for classification. 

The remaining structure of this paper is as follows. Section \ref{Methodology} describes the datasets used, the proposed framework, and the model training and testing. Section \ref{Experiments} presents the experimental setup and results for the training dataset and testing datasets. Finally, the discussion, conclusions, and future work are given in Section \ref{conclusion}.

\section{Methodology}
\label{Methodology}
An overview of the SupWMA framework is given in Fig.~\ref{fig_overview}. In the rest of this section, first we introduce relevant datasets (Section \ref{dMRIDatasets}), then we present our two-stage framework and contrastive learning methodologies (Sections \ref{TwoStageFramework}-\ref{SCL}), and finally we describe training and testing procedures (Sections \ref{TwoStageTraining}-\ref{SWMInference}).

\subsection{dMRI datasets and tractography}
\label{dMRIDatasets}

\begin{table*}[!t]
	\footnotesize
	\caption{\textmd{Demographics and dMRI acquisition of the six independently acquired datasets for experimental evaluation.}}
	\centering
	\begin{threeparttable}
		\begin{tabular}{| L{1.1cm} | L{1.cm} | L{1.7cm} | L{1.cm} | L{1.9cm} | L{5cm} |}
			\hline
			Dataset & Number & Age & Gender & Healthy/Disease & dMRI data \\
			\hline
            HCP & 100 & 22 to 35 y \newline (29.0 $\pm$ 3.5) & 54 F\newline 46 M & 100 H & b=3000 s/mm$^2$; 108 directions; \newline TE/TR=89/5520 ms; resolution=1.25 mm$^3$\\
			\hline
            dHCP & 40 & 1 to 27 d\newline(6.30 $\pm$ 7.47)
             &  15 F\newline 25 M & 40 H & b=400/1000/2600 s/mm$^2$; 300 directions; \newline TE/TR=90/3800 ms; resolution=1.5 mm$^3$ \\
			\hline
            ABCD & 50& 9 to 11 y \newline (10.02 $\pm$ 0.69) & 25 F \newline 25 M & 50 H & b = 3000 s/mm$^2$; 96 directions; \newline TE/TR = 88/4100 ms; resolution = 1.7 mm$^3$ \\
			\hline
            CNP & 50 & 21 to 50 y \newline (32.36 $\pm$ 8.65) & 23 F \newline 27 M & 14 H \newline 12 ADHD \newline 12 BD \newline 12 SZ & b = 1000 s/mm$^2$; 64 directions; \newline TE/TR = 93/9000 ms; resolution = 2~mm$^3$ \\
			\hline
            PPMI & 50 & 45 to 80 y \newline (62.52 $\pm$ 6.95) & 13 F \newline 37 M & 25 H \newline 25 PD & b = 1000 s/mm$^2$; 64 directions; \newline TE/TR = 88/7600 ms; resolution = 2 mm$^3$ \\
			\hline
            BTP & 39 & 23 to 82 y \newline (48.9 $\pm$ 15.3) & 16 F \newline 23 M & 39 BTP & b=2000 s/mm$^2$; 31 directions; \newline TE/TR=98/12700ms; resolution=2.3 mm$^3$ \\
			\hline
		\end{tabular}
		\begin{tablenotes}
			\item \textit{Abbreviations}: Dataset: HCP - Human Connectome Project \citep{Van_Essen2013-ne} (These subjects are different from those used in the atlas); dHCP - Developing Human Connectome Project \citep{Edwards2022-vr}; ABCD - Adolescent Brain Cognitive Development \citep{Volkow2018-og}; CNP - Consortium for Neuropsychiatric Phenomics \citep{Poldrack2016-zs}; PPMI - Parkinson’s Progression Markers Initiative \citep{Parkinson_Progression_Marker_Initiative2011-oz}; BTP - Brain Tumor Patient. Age: d - day; y - year. Gender: F - female; M - male. Healthy/Disease: H - healthy; ADHD - attention-deficit/hyperactivity disorder; BP - bipolar disorder; SZ - schizophrenia; PD - Parkinson’s disease; BTP - brain tumor patient.
		\end{tablenotes}
	\end{threeparttable}
	\label{tab_demo_summary}
\end{table*}

\textbf{Training dataset:}
A high-quality large-scale tractography dataset with 1 million labeled streamlines was used for model training and validation. This dataset was derived from an anatomically curated white matter tractography atlas \citep{Zhang2018-jx}, which was provided by the O’Donnell Research Group (ORG) and is available to access online\footnote{\href{https://github.com/SlicerDMRI/ORG-Atlases}{https://github.com/SlicerDMRI/ORG-Atlases}}. The ORG atlas has been previously used for DWM streamline classification with deep learning \citep{Zhang2020-vm}. In brief, the atlas was generated by creating dense tractography maps \citep{Malcolm2010-mk} of 100 young healthy adults in the Human Connectome Project (HCP) \citep{Van_Essen2013-ne} and applying a fiber clustering method \citep{ODonnell2007-sa,ODonnell2012-ol,Zhang2018-jx} to group streamlines across subjects according to their similarity in shape and location. More specifically, this method performed spectral embedding of streamlines to learn a high-dimensional space where each streamline had a unique representation. In this space, a k-means clustering was performed to group the streamlines into multiple clusters. Clusters do not overlap in the high-dimensional spectral space, enabling unambiguous cluster definitions. For each cluster, a data-driven process was then applied to identify apparent outlier streamlines, defined as those that had a fiber similarity over 2 standard deviations from the cluster's mean fiber similarity (as in \citep{ODonnell2017-sq,Zhang2018-gd,Zhang2018-jx}). In this way, every cluster in the ORG atlas obtained a corresponding group of outlier streamlines. A neuroanatomist then annotated anatomical labels for clusters in the atlas. Atlas clusters are bilateral (clusters have similar shapes and cortical projections across both left and right hemispheres). In comparison with other publicly available SWM atlases \citep{Guevara2020-da}, the ORG atlas provides comprehensive coverage of the SWM and a fine parcellation scale. The ORG atlas also includes labeled outlier streamlines \citep{Zhang2020-vm}. The stage-one training dataset ($D_1$) includes all 1 million streamlines, where each streamline is labeled as either SWM or DWM. The stage-two training dataset ($D_2$) includes only SWM streamlines (237896 streamlines), where streamline labels include 198 SWM cluster classes and 198 corresponding SWM outlier categories. Each streamline has one label. Note that SWM streamlines in the ORG atlas are long- and medium-range SWM connections (over 40~mm in length).

\textbf{Testing datasets:} For experimental evaluation, we used data from 329 subjects from six independently acquired datasets (testing datasets) with different imaging protocols across the lifespan (neonates, children, young adults and older adults, ranging in age from 1 day to 82 years) and health conditions (healthy control, Parkinson’s patients, neuropsychiatric disorders, and neurosurgical patients with brain tumors). Tractography streamlines in these testing  datasets do not have labels. Table \ref{tab_demo_summary} presents an overview of the demographics and acquisition protocols for the six testing datasets. HCP \citep{Van_Essen2013-ne}, dHCP~\citep{Edwards2022-vr}, ABCD \citep{Volkow2018-og}, CNP \citep{Poldrack2016-zs} and PPMI \citep{Parkinson_Progression_Marker_Initiative2011-oz} datasets are publicly available. Image data from the BTP dataset was acquired at Brigham and Women’s Hospital. The Partners Healthcare Institutional Review Board approved the data usage, and all participants gave informed consent prior to scanning. 

Whole-brain tractography was performed on all testing datasets using the two-tensor Unscented Kalman Filter (UKF)\footnote{\href{https://github.com/pnlbwh/ukftractography}{https://github.com/pnlbwh/ukftractography}} method \citep{Malcolm2010-mk,Reddy2016-ko}. This method works robustly across acquisitions and different populations including neonates, children, adults and patients with peritumoral edema \citep{Chen2015-fy,Zhang2018-jx,Zhang2020-vm}. To perform tractography we used the same parameter settings used for generating tractography in the ORG atlas \citep{Zhang2018-jx}. In brief, tractography was seeded in all voxels within the brain mask where fractional anisotropy (FA) was greater than 0.1. Tracking stopped where the FA value fell below 0.08 or the normalized mean signal (the sum of the normalized signal across all gradient directions) fell below 0.06. Streamlines that were longer than 40~mm were retained \citep{Guevara2012-ex,Jin2014-wv,Zhang2018-jx}. On average across testing datasets, each subject’s tractography data contained approximately 0.65 million streamlines, with an average streamline length of about 92 mm. The number of points along a streamline depended on the streamline’s length (points were stored approximately every 0.9 mm along streamlines). Tractography quality control was performed using \textit{whitematteranalysis} software in SlicerDMRI \citep{Norton2017-nz,Zhang2020-cv}. Tractography data were registered into the space of the training dataset using an affine transform produced by registering the b0 image of each subject to the mean T2-weighted image of the ORG atlas using 3D Slicer\footnote{\href{https://www.slicer.org}{https://www.slicer.org}} \citep{Fedorov2012-ma}.

\subsection{Two-stage framework for SWM parcellation}
\label{TwoStageFramework}
The goal of our two-stage framework (Fig. \ref{fig_overview}(b)) is to decompose the complicated SWM parcellation problem into two sub-problems: a binary classification of SWM and DWM streamlines (stage one), and a multi-class classification of 198 SWM clusters and 198 SWM outlier categories (stage two). Stage one handles the classification of SWM vs DWM streamlines (which is non-trivial, as these categories cannot be easily distinguished by a simple threshold on streamline length as shown in Supplementary Fig. S1). Stage two handles the actual parcellation of the SWM and enables removal of streamline outliers from each cluster. Multi-stage frameworks that divide the target task into subtasks have also been applied successfully in many other medical imaging tasks \citep{Wu2020-va,Al-masni2020-oy,Panda2021-tt,Zhang2017-ei,Jain2020-cb}. In our case, the two stages have the same network structure for point cloud classification (Section \ref{PointCloudNetwork}) but different training data and procedures (Section \ref{TwoStageTraining}). The two-stage inference process for SWM parcellation is described in Section \ref{SWMInference}. 

Stage one takes whole-brain tractography streamlines as input, and outputs streamlines classified as SWM and DWM. In this way, stage one filters out DWM streamlines, which are not of interest for SWM parcellation. Stage one is trained with cross-entropy (CE) loss on the training dataset $D_1$. The task in stage one is relatively straightforward: stage one differentiates SWM and DWM streamlines, which have relatively large distances in the latent global feature space due to the spatial differences between SWM and DWM anatomy.

Stage two takes SWM streamlines as input, and it outputs streamlines classified into 198 SWM clusters and outliers. Stage two is trained with both cross-entropy loss and supervised contrastive loss \citep{Khosla2020-mx} on the training dataset $D_2$ (Section \ref{SCL}). The task in stage two is more challenging than that of stage one, because plausible streamlines from different clusters, as well as outlier streamlines, can have similar trajectories (Fig. \ref{fig_overview}(a)). Therefore, we incorporate supervised contrastive learning in this stage to enhance the feature learning (Section \ref{SCL}). 

\subsection{Point cloud network structure}
\label{PointCloudNetwork}
\begin{figure}[t] 
\centering
\includegraphics[width=0.48\textwidth]{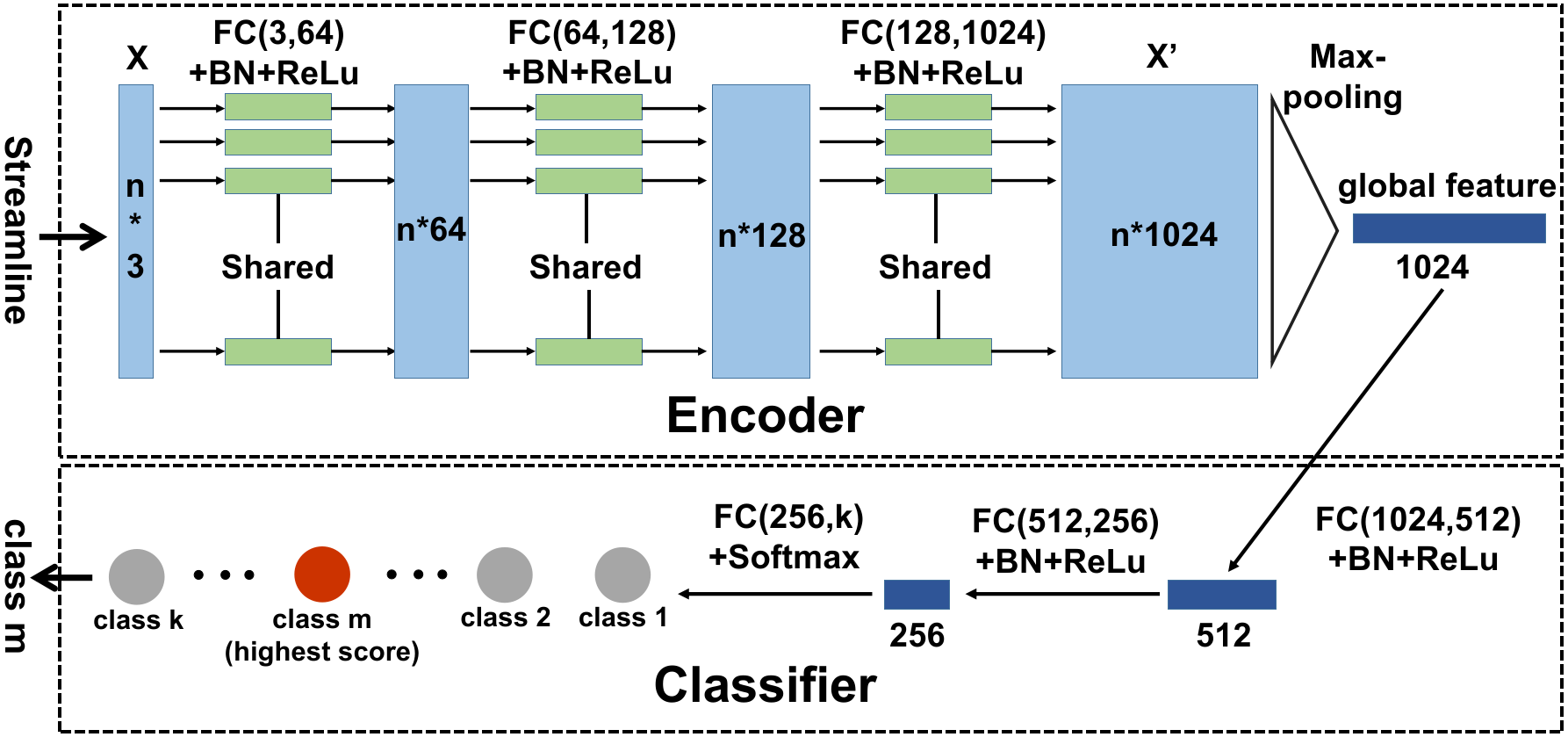}
\caption{\textmd{Diagram of the point-cloud-based network architecture.}}
\label{fig_network}
\end{figure}

The proposed network (Fig. \ref{fig_network}) is based on PointNet \citep{Charles2017-nv}, which is widely used for point cloud classification and segmentation \citep{Guo2020-ql}. PointNet includes a shared multi-layer perceptron (MLP), a symmetric aggregation function, and fully connected (FC) layers, as well as data-dependent transformation nets. The transformation nets are designed to perform affine transformation of input point clouds into a canonical space \citep{Charles2017-nv}. However, in our framework we remove the transformation nets to preserve important information about the spatial position of streamlines in the brain while improving the computational speed of the network.

The proposed network structure (Fig. \ref{fig_network}) is used in both stage one and stage two. Our network includes two parts: the encoder \textit{Enc(·)} that extracts a non-linear global feature for each streamline and the classifier \textit{Cla(·)} that predicts the streamline label using the extracted global feature in latent space. Specifically, for \textit{Enc(·)}, the input is the RAS (Right, Anterior, Superior) coordinates of streamline points, denoted as $X=(x_1, x_2, ..., x_n)$, where ${x_i}$ is the 3-D coordinate of point $i$. Therefore, the dimension of $X$ is $n\times$3 (where $n$ = 15, an appropriate number for streamline representations \citep{Zhang2018-jx,ODonnell2007-sa,Zhang2020-vm}). Each point $x_i$ is individually encoded by a shared MLP of three layers that have 64, 128, and 1024 output dimensions, respectively. Each layer includes batch normalization and rectified linear units (ReLU) activation. Therefore, an output $X’$ with $n\times$1024 dimensions is generated, where $X’ = (MLP(x_1), MLP(x_2), ..., MLP(x_n))$. Then, a symmetric function (max-pooling \citep{Charles2017-nv}) aggregates the encoded features $X’$ to form a 1024-dimension global shape feature $g$ of the streamline. $g$ is invariant to the order of points along a streamline, so that streamlines with equivalent forward and reverse point orderings are allowed to have the same global shape feature (representation). Finally, the \textit{Cla(·)}, consisting of three ordinary FC layers with sizes of 512, 256, and $k$ (number of output classes), is used for streamline class prediction. The first and second FC layers are both followed by batch normalization and ReLU activation. The third FC layer followed by a softmax layer is used to output the prediction result (the class with the highest score).

\subsection{Supervised contrastive learning}
\label{SCL}
Supervised contrastive learning (SCL) \citep{Khosla2020-mx} extends self-supervised contrastive learning \citep{Chen2020-pz} to a fully-supervised mode by proposing a supervised contrastive loss, which aims to pull global features (outputs of \textit{Enc(·)}) with the same class label closer in the latent space and push apart global features with different class labels. SCL has been shown to successfully improve the performance of supervised learning in computer vision \citep{Schiffer2021-vl,Khosla2020-mx,Kopuklu2021-vt,Zhong2021-od} and natural language processing \citep{Han2021-ke,Gunel2020-zq,Huang2021-fl} tasks.

\begin{figure}[t] 
\centering
\includegraphics[width=0.48\textwidth]{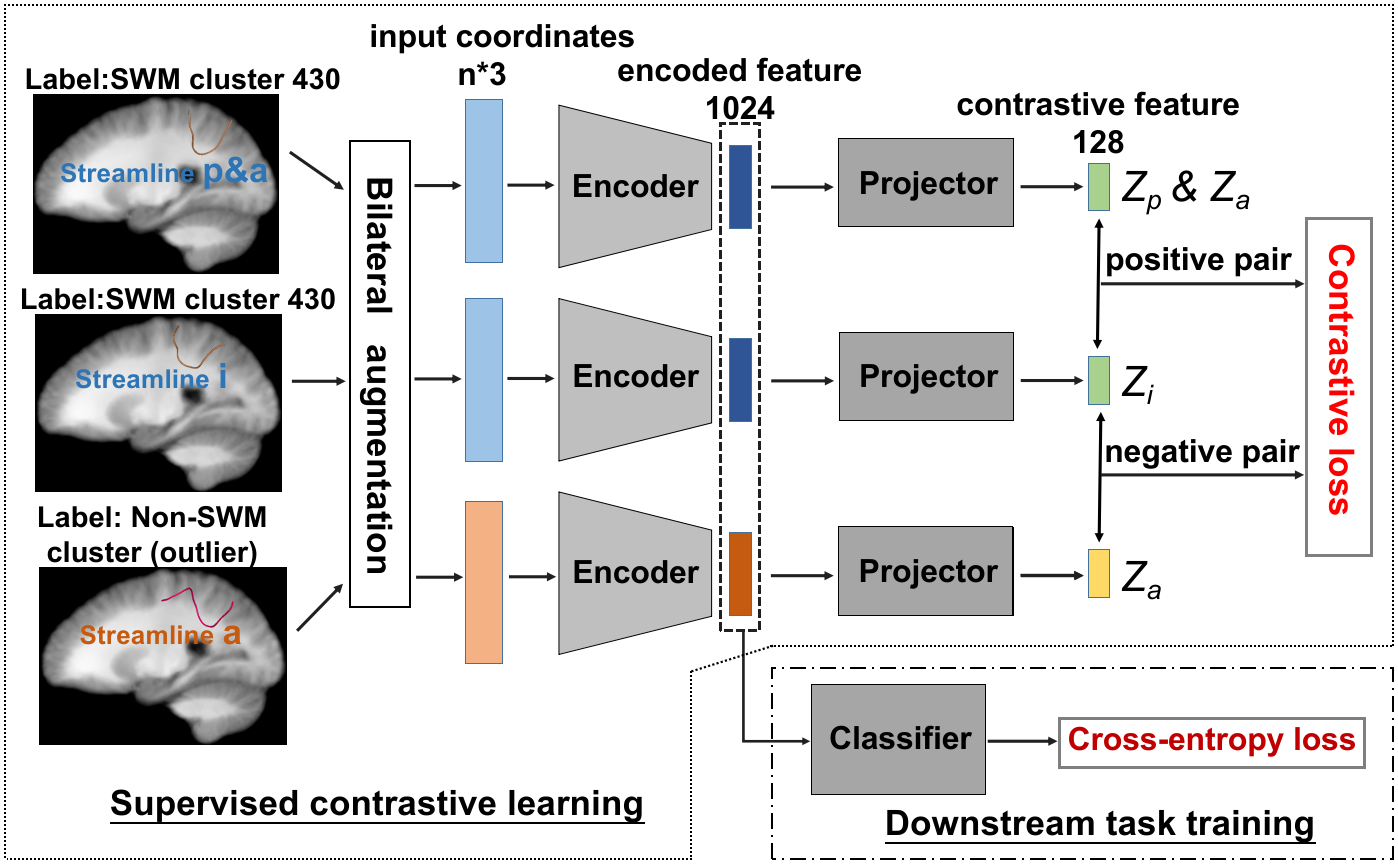}
\caption{\textmd{Training procedure in stage two: the process of supervised contrastive learning and downstream task training}}
\label{fig_scl}
\end{figure}

In the proposed framework, we employ SCL in stage two to obtain more distinguishable global features for SWM clusters and outliers. In SCL training for \textit{Enc(·)} (Fig. \ref{fig_scl}), a projector head \textit{Proj(·)} \citep{Khosla2020-mx,Chen2020-pz} is added on top of \textit{Enc(·)}. \textit{Proj(·)} has two additional FC layers of sizes 1024 and 128 followed by a normalization layer. Therefore, the contrastive feature \textit{z} of input \textit{X} for calculating contrastive loss is formed as $z=Proj(g)=Proj(Enc(X))$. \textit{Proj(·)} may retain more instance-specific streamline information in the global feature $g$ to benefit downstream tasks \citep{Chen2020-pz}.

The supervised contrastive loss is defined as:
\begin{equation*}
    \mathcal{L}=\sum_{i \in I}\mathcal{L}_{i}=\sum_{i \in I} \frac{-1}{|P(i)|}  \sum_{p \in P(i)} \log \frac{\exp ( z_{i} \cdot z_{p} / \tau )}{\sum_{a \in A(i)} \exp ( z_{i} \cdot z_{a} / \tau )}
\end{equation*}
where $I$ is the streamline set in a training batch ($i\in I \equiv \{1,..., M\}$); $P(i)$ is the streamline set that has the same class label as streamline $i$ ($p \in P(i)$); $A(i)$ is the set of all other streamlines in $I$ except for streamline $i$ ($a \in A(i) \equiv I \backslash \{i\}$); $z_i$, $z_p$ and $z_a$ are contrastive features obtained from \textit{Proj(·)} for streamlines $i$, $p$ and $a$; $\tau$ (temperature) is a pre-defined hyperparameter set to be 0.1 as suggested in \citep{Chen2020-pz}.

Since our clusters in training are bilateral, with similar shapes and cortical projections across both left and right hemispheres (see details in Supplementary Material 2), we employed a data augmentation technique, called bilateral augmentation \citep{ODonnell2007-sa,Zhang2018-gd}, for improving the performance of SCL. Data augmentation techniques have been shown to be helpful in contrastive learning \citep{Chen2020-pz,Khosla2020-mx}, where feature representations of augmented data from meaningful transformations are encouraged to be invariant to representations of the original sample. Bilateral augmentation generates a symmetric streamline in the other hemisphere, and the generated streamline has the same class label as the original streamline. Therefore, it encourages SCL to obtain the same representation for symmetric streamlines. In addition, the number of streamlines is doubled by applying bilateral augmentation in each training batch. 

\subsection{Two-stage model training}
\label{TwoStageTraining}
In stage one, \textit{Enc(·)} and \textit{Cla(·)} are trained together with CE loss. The learning rate is 0.001 (selected from a grid search of 0.01, 0.001, and 0.0001) and batch size is 1024 (selected from a grid search of 256, 512, 1024, 2048). Relatively large batch sizes are preferred since the number of streamlines (1 million) in our training dataset is large. Therefore, we chose large numbers for the grid search of batch sizes. In stage two, training includes two phases: contrastive learning and downstream learning \citep{Khosla2020-mx,Chen2020-pz}. In the contrastive learning phase, \textit{Enc(·)} and \textit{Proj(·)} are trained with supervised contrastive loss. The learning rate is 0.01 (selected from a grid search of 0.01, 0.001, and 0.0001), and the batch size is 3072 (selected from fine-tuning based on suggestions in \citep{Khosla2020-mx,Chen2020-pz}). In the downstream learning phase, the parameters of \textit{Enc(·)} are frozen as in \citep{Khosla2020-mx}, and \textit{Proj(·)} is untouched. \textit{Cla(·)} takes $g$ (the output of \textit{Enc(·)}) as the input and is trained with CE loss for predicting streamline labels. The learning rate is 0.001, and the batch size is 1024. The selection process of learning rate and batch sizes for \textit{Cla(·)} is the same as stage one. All training processes utilize Adam \citep{Kingma2014-hv} as the optimizer with no weight decay (selected from a grid search of 0, 0.0001, 0.00001). Training and validation were performed with Pytorch (v1.7.0) on a NVIDIA GeForce RTX 2080 Ti GPU machine.

\subsection{Inference for SWM parcellation}
\label{SWMInference}
For inference, the trained two-stage model is applied for SWM parcellation of unlabeled tractography data. The two-stage prediction process with trained networks works as follows. In stage one, streamlines with predicted labels of DWM are filtered out, and streamlines with predicted labels of SWM are retained for input to stage two. In stage two, the model takes these predicted SWM streamlines from stage one and classifies them into 198 SWM clusters and SWM outliers. Therefore, the final overall SWM parcellation result includes 198 SWM clusters and one non-SWM cluster (which contains SWM outliers and DWM streamlines).

\section{Experiments and results}
\label{Experiments}
In this section, we first introduce experiments on the training dataset where streamline labels are available (Section \ref{ExperimentTrainingData}), followed by the experiments on independently acquired testing datasets (Section \ref{ExperimentTestingData}).

\subsection{Experiments on the training dataset}
\label{ExperimentTrainingData}
We conducted three experiments using 5-fold cross-validation on the training dataset. First, we compared SupWMA to two deep-learning-based state-of-the-art (SOTA) methods. Second, we assessed the performance of ablated versions of our proposed SupWMA framework. Third, we analyzed the importance of different points along a streamline, e.g., endpoints or center points, for streamline classification within our proposed SupWMA framework. 

\subsubsection{Comparison with SOTA deep learning methods}
We selected two SOTA deep-learning-based tractography parcellation methods (DeepWMA\footnote{\href{https://github.com/zhangfanmark/DeepWMA}{github.com/zhangfanmark/DeepWMA}} \citep{Zhang2020-vm} and DCNN+CL+ATT\footnote{\href{https://github.com/HaotianMXu/Brain-fiber-classification-using-CNNs}{github.com/HaotianMXu/Brain-fiber-classification-using-CNNs}} \citep{Xu2019-aa}) for comparison to the proposed SupWMA method. 

Deep white matter analysis (DeepWMA) \citep{Zhang2020-vm} is designed for DWM parcellation using a Convolutional Neural Network (CNN) and streamline spatial coordinate features. DeepWMA has eight CNN layers and three FC layers. It uses a “FiberMap” input feature that converts spatial coordinates of streamlines into 3-channel images that enable streamlines with forward and backward point orders to have nearly equivalent representations. DeepWMA achieves consistent DWM parcellation results across populations \citep{Zhang2020-vm}. 

DCNN+CL+ATT \citep{Xu2019-aa} is also designed for DWM parcellation. DCNN (Deep CNN, inspired by \citep{He2015-nc} and adapted from \citep{Xu2018-rk}) is employed with soft spatial attention (ATT) modules \citep{Xu2015-ns}. During training, this method employs two losses: focal loss \citep{Lin2017-or} to help in unbalanced datasets and center loss (CL) \citep{Wen2016-zn} to assist the network to obtain better streamline representations. This framework obtains satisfactory accuracy for predicting functionally important white matter pathways that should be protected in the surgery of epilepsy patients \citep{Xu2019-aa}.

For both of these approaches, we trained their networks and tuned hyperparameters based on the suggested settings in their papers and released codes. For evaluation, we computed the accuracy and macro F1-score metrics, which have been widely used for the evaluation of tractography parcellation \citep{Liu2019-jj,Ngattai_Lam2018-ne,Zhang2020-vm,Xu2019-aa}. For each cross-validation fold, the accuracy of streamline classification was calculated, and the mean and standard deviation of the macro F1-score across 199 streamline classes were also reported. Finally, the average of each metric across the five folds was computed and is presented in Table~\ref{tab_sota_compare}. We also include the floating point operations (FLOPs), which measure the number of required operations performed for model inference \citep{Charles2017-nv}, for evaluating the efficiency of each method.


\begin{table}[htbp]
\footnotesize
\caption{\textmd{Quantitative comparisons of SOTA deep-learning-based methods on the training dataset.}}
\centering
\begin{tabular}{|c|c|c|c|}
\hline
 Methods &  Accuracy & F1-score  &  FLOPs/streamline \\
\hline
 DeepWMA & 92.74\% & 73.35$\pm$7.93\% & 40.87M \\
 DCNN+CL+ATT & 95.31\% & 83.01$\pm$4.09\% & 36.82M \\
SupWMA & \textbf{96.79\%} & \textbf{88.79$\pm$2.91\%} & \textbf{5.68M} \\
\hline
\end{tabular}
\label{tab_sota_compare}
\end{table}

Compared to the two SOTA methods (Table~\ref{tab_sota_compare}), SupWMA achieves the highest mean accuracy (96.79\%) and macro F1-score (88.79\%) with the lowest standard deviation. SupWMA outperforms DeepWMA and DCNN+CL+ATT by 4.05\% and 1.48\% in accuracy. Also, the F1-score of SupWMA is 15.44\% and 5.78\% higher than DeepWMA and DCNN+CL+ATT, respectively. Furthermore, the FLOPs of SupWMA are much lower than other SOTA methods. Overall, these results demonstrate the benefits of our novel deep learning framework to enable accurate and efficient SWM parcellation.


\begin{table*}[htbp]
\footnotesize
\caption{\textmd{Ablation experiments on the training dataset.}}
\centering
\begin{tabular}{|c|c|c|c|c|}
\hline
Methods & Accuracy & F1-score  & FLOPs/streamline & Inference time/streamline \\
\hline
 PointNet\textsubscript{orignial} & 96.11\% & 86.48$\pm$3.47\% & 9.58M & 13.07ms \\
 PointNet\textsubscript{two-stage} & 96.39\% & 87.15$\pm$3.37\% & 19.16M & 15.41ms \\
SupWMA\textsubscript{two-stage} & 96.48\% & 87.47$\pm$3.18\% & 5.68M & 6.82ms\\
SupWMA\textsubscript{two-stage+SCL} & \textbf{96.79\%} & \textbf{88.79$\pm$2.91\%} & \textbf{5.68M} & \textbf{6.77ms}\\
\hline
\end{tabular}
\label{tab_ablation}
\end{table*}

\begin{table*}[htbp]
\footnotesize
\caption{\textmd{Cluster identification rates across methods for six testing datasets.}}
\centering
\begin{tabular}{|c|c|c|c|c|c|c|}
\hline
Methods & HCP & dHCP  & ABCD & CNP & PPMI & BTP \\
\hline
WMA & 97.86$\pm$2.37\%  & 60.13$\pm$17.19\%  & 89.92$\pm$5.14\% & 94.58$\pm$5.12\% & 96.35$\pm$3.12\% & 83.88$\pm$9.85\% \\
DeepWMA & 98.49$\pm$1.82\% & 64.27$\pm$15.36\% & 89.55$\pm$3.83\% & 95.63$\pm$2.50\% & 
96.86$\pm$2.15\% & 85.02$\pm$7.63\%\\
DCNN+CL+ATT  & 99.02$\pm$1.64\% & 65.35$\pm$16.84\% & 90.85$\pm$3.88\% & 96.23$\pm$2.74\% & 
96.97$\pm$2.39\% & 85.21$\pm$8.15\% \\
SupWMA & \textbf{99.28$\pm$1.41\%} & \textbf{68.29$\pm$17.24\%} & \textbf{93.61$\pm$3.31\%} & \textbf{96.76$\pm$2.66\%} & \textbf{97.35$\pm$2.39\%} & \textbf{86.62$\pm$7.92\%} \\
\hline
\end{tabular}
\label{tab_CIR}
\end{table*}

\subsubsection{Ablation study}

To analyze the effectiveness of our framework design, we performed an ablation study including the following: PointNet\textsubscript{orignial} that is the original PointNet implementation (with transformation nets equipped),  PointNet\textsubscript{two-stage} that employs the original PointNet with a two-stage framework, SupWMA\textsubscript{two-stage} that removes transformation nets from SupWMA\textsubscript{two-stage}, and the proposed SupWMA\textsubscript{two-stage+SCL} that uses SCL and the two-stage SWM parcellation framework. In addition to accuracy, F1-score, and FLOPs per streamline, we also calculated the inference time per streamline in 5-fold cross validation.

The ablation study (Table~\ref{tab_ablation}) shows that two-stage design in PointNet\textsubscript{two-stage} increases accuracy and macro F1-score compared to PointNet\textsubscript{orignial}. Although the FLOPs per streamline are doubled using PointNet\textsubscript{two-stage}, the computational speed for SWM parcellation is not greatly reduced due to the two-stage parcellation framework that first removes a large number of DWM streamlines. Inference time per streamline slightly increases from 13.07ms for PointNet\textsubscript{orignial} to 15.41ms for PointNet\textsubscript{two-stage} with a two-stage parcellation framework. SupWMA\textsubscript{two-stage} further improves the model performance and reduces the FLOPs as well as the inference time compared to PointNet\textsubscript{two-stage}, showing the advantages of removing the transformation nets for our application. Finally, the proposed SupWMA\textsubscript{two-stage+SCL} achieves the best performance, i.e., the highest accuracy, the highest F1-score, the lowest FLOPs, and the shortest inference time. These demonstrate the benefits of our network structure, two-stage parcellation framework and designed SCL that can extract highly discriminative global features of streamlines for parcellation. 

\subsubsection{Critical streamline point indices for SWM classification}
To study how our model classifies SWM streamlines, we explore the importance scores of point indices \citep{Charles2017-nv} for SWM streamline classification. Importance scores of point indices are quantified by the max pooling operation with our trained model. Since we implement 5-fold cross-validation for evaluating performance, five SupWMA trained models with the best average F1-score on five folds are selected. We input the validation data (20\% of the entire dataset) into its corresponding trained model for each fold to get importance scores for all streamline points in the training dataset. Importance scores are obtained using the stage two model to focus this experiment on SWM (after removal of DWM). Importance scores for each point index are obtained from the max-pooling operation in the trained encoder. To be specific, we have 15 points for each streamline, and each point is encoded as a 1024-dimension (d) feature vector. For each dimension in the 1024-d feature vector, the max-pooling operation returns the maximum value and its point index out of 15 point values. Therefore, we can obtain the number of times ($N_{max}$) a point index is returned by the max-pooling operation. The possible $N_{max}$ for each point index ranges from 0 to 1024. The higher the $N_{max}$, the more important the point is. 

\begin{figure}[b] 
\centering
\includegraphics[width=0.48\textwidth]{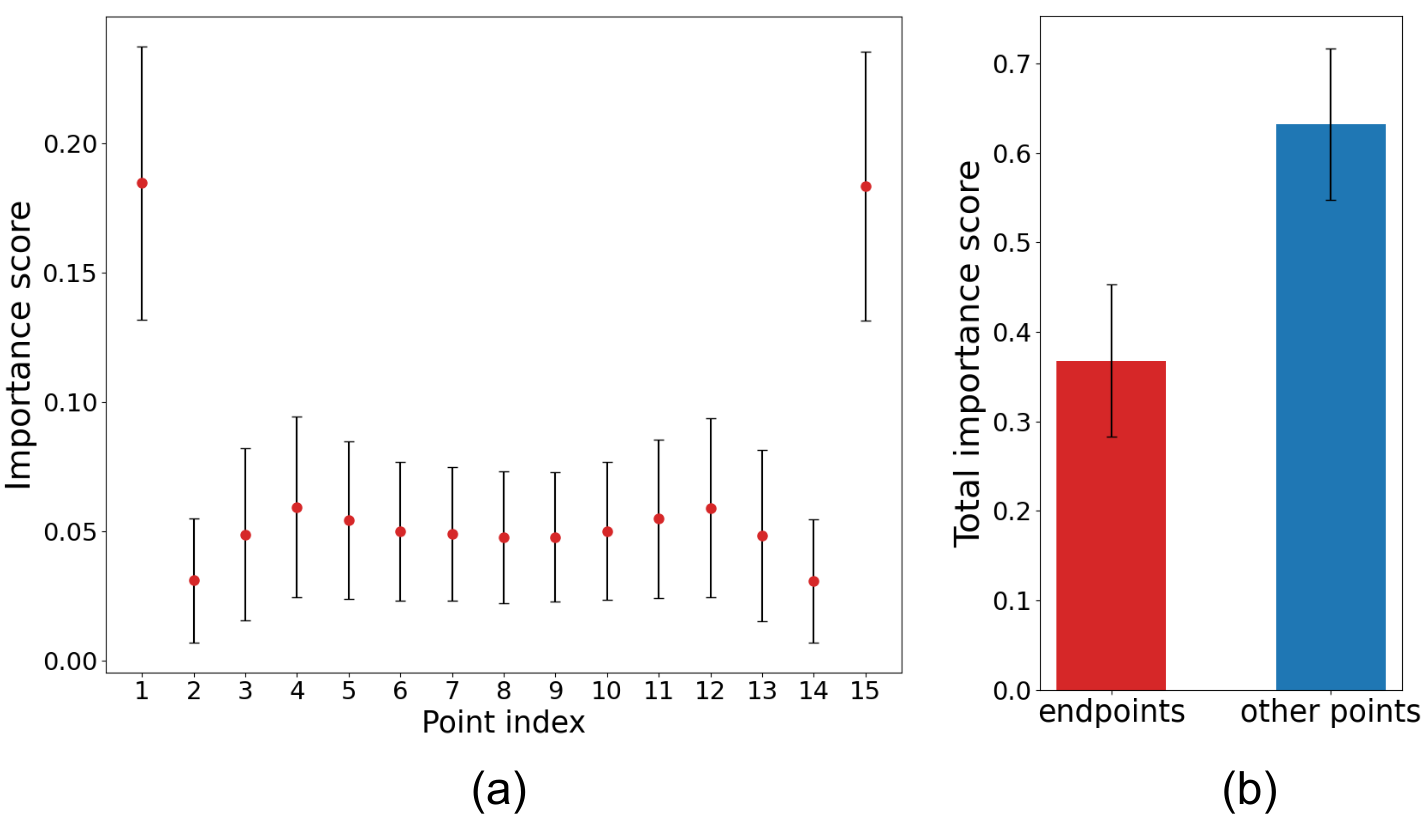}
\caption{\textmd{(a) The average importance score and its standard deviation for each point index across streamlines. (b) The total importance scores of the two streamline endpoints and all of the other streamline points.}}
\label{fig_critical_points}
\end{figure}

Fig.~\ref{fig_critical_points}(a) presents the average importance score (normalized $N_{max}$) and its standard deviation for each point index across streamlines in the entire training dataset. It can be seen that the two streamline endpoints are the most critical points for classification of SWM streamlines. Fig.~\ref{fig_critical_points}(b) shows the total importance scores of the two endpoints (red bar) and the other points (blue bar). While the two endpoints are the most informative, the other points along the streamline also provide important information for SWM classification.

\begin{figure*}[t] 
\centering
\includegraphics[width=1.0\textwidth]{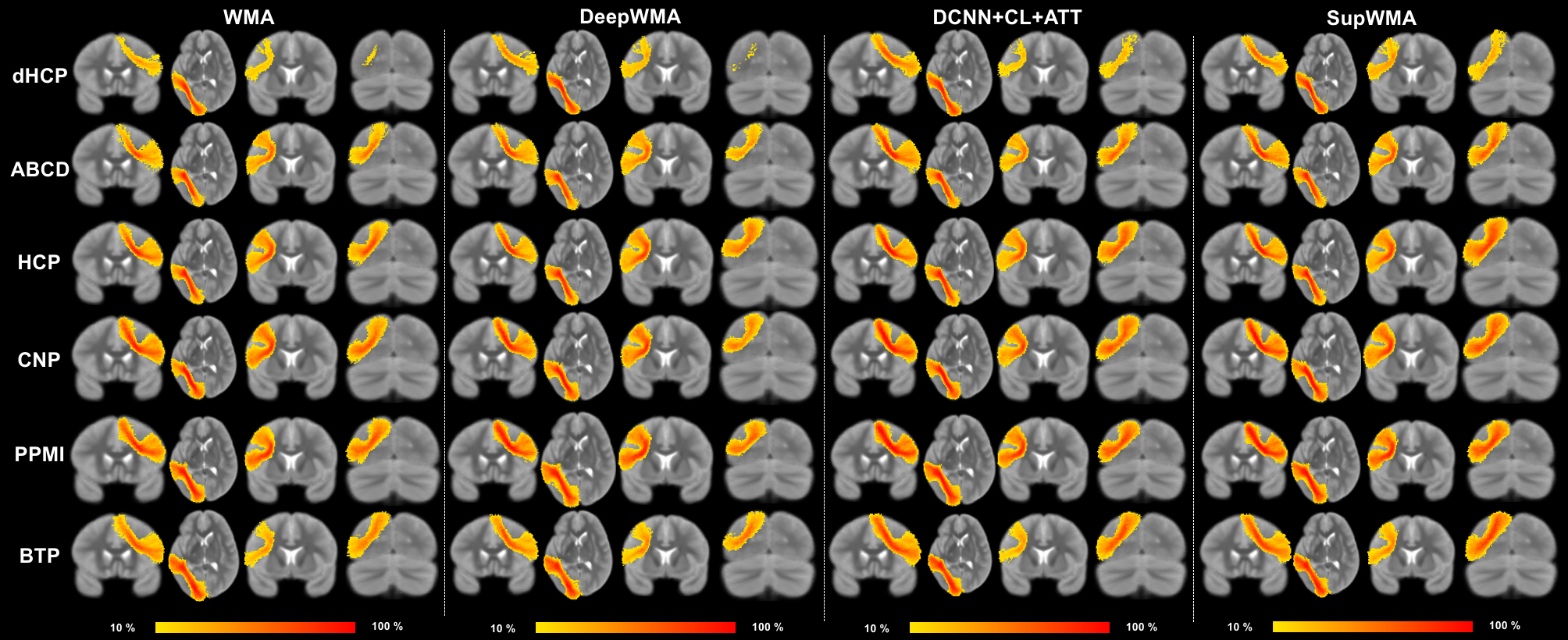}
\caption{\textmd{Visual comparison of the voxel-based population cluster heatmaps of four example clusters from results of WMA, DeepWMA, DCNN+CL+ATT, and SupWMA. Heatmaps depict the percentage of subjects that have streamlines passing through each voxel (see colorbar). The population mean T2 image of the ORG atlas is selected as the background image.~Note that as these heatmaps show each population as a whole, the BTP heatmaps are not highly affected by patient-specific individual tumor locations.}}
\label{fig_heatmap}
\end{figure*}

\begin{table*}[htbp]
\footnotesize
\caption{\textmd{Cluster distance to atlas (unit: mm) across methods for six testing datasets.}}
\centering
\begin{tabular}{|c|c|c|c|c|c|c|}
\hline
Methods & HCP & dHCP  & ABCD & CNP & PPMI & BTP \\
\hline
WMA & 4.096$\pm$0.632  & 5.462$\pm$1.223  & 4.457$\pm$0.735 & 4.630$\pm$0.846 & 4.414$\pm$0.760 & 4.619$\pm$0.912 \\
DeepWMA & 4.035$\pm$0.627 & 5.544$\pm$2.182 & 4.275$\pm$0.718 & 4.350$\pm$0.725 & 4.273$\pm$0.688 & 4.576$\pm$1.141\\
DCNN+CL+ATT  & \textbf{3.882$\pm$0.518} & 4.841$\pm$2.369 & 4.141$\pm$0.893 & \textbf{4.179$\pm$0.667} & \textbf{4.097$\pm$0.543} & 4.361$\pm$2.220 \\
SupWMA & 3.902$\pm$0.530 & \textbf{4.785$\pm$0.813} & \textbf{4.133$\pm$0.578} & 4.190$\pm$0.596 & 4.123$\pm$0.574 & \textbf{4.303$\pm$0.730} \\
\hline
\end{tabular}
\label{tab_CDA}
\end{table*}

\begin{table*}[ht]
\footnotesize
\caption{\textmd{Inter-subject parcellation variability across methods for six testing datasets.}}
\centering
\begin{tabular}{|c|c|c|c|c|c|c|}
\hline
Methods & HCP & dHCP  & ABCD & CNP & PPMI & BTP \\
\hline
WMA & 0.7046$\pm$0.1912  & 1.7164$\pm$0.7683  & 0.8659$\pm$0.2826 & 0.8633$\pm$0.2902 & 
0.8081$\pm$0.2458 & 1.1275$\pm$0.3560\\
DeepWMA & 0.6496$\pm$0.1732 & 1.5832$\pm$0.6934 & 0.8169$\pm$0.3120 & 0.8170$\pm$0.3313 & 
0.7750$\pm$0.2652 & 1.0996$\pm$0.4066\\
DCNN+CL+ATT  & 0.6424$\pm$0.1500 & 1.5372$\pm$0.6062 & 0.8106$\pm$0.2762 & 0.8143$\pm$0.2740 & 
0.7748$\pm$0.2439 & 1.0901$\pm$0.3570 \\
SupWMA & \textbf{0.6323$\pm$0.1399} & \textbf{1.4759$\pm$0.5422} & \textbf{0.7799$\pm$0.2398} & \textbf{0.8008$\pm$0.2669} & \textbf{0.7636$\pm$0.2334} & \textbf{1.0626$\pm$0.3453} \\
\hline
\end{tabular}
\label{tab_ISPV}
\end{table*}

\subsection{Experiments on independently acquired testing datasets}
\label{ExperimentTestingData}
We also performed experiments on six independently acquired testing datasets (Table \ref{tab_demo_summary}), where streamline labels were not available. In addition to the two deep-learning-based SOTA methods (DeepWMA and DCNN+CL+ATT, introduced in Section~\ref{ExperimentTrainingData}), we also compared \textit{whitematteranalysis} (WMA)\footnote{\href{https://github.com/SlicerDMRI/whitematteranalysis}{https://github.com/SlicerDMRI/whitematteranalysis}} \citep{ODonnell2007-sa,Zhang2018-jx,ODonnell2012-ol}. WMA performs spectral clustering \citep{ODonnell2007-sa} and entropy-based registration \citep{ODonnell2012-ol} for white matter parcellation. Several deep learning papers \citep{Wasserthal2019-ha,Zhang2020-vm,Chen2021-dm} have employed WMA as a comparison method. We note that WMA performs tractography parcellation by applying the same anatomically curated atlas (see \citep{Zhang2018-jx} for details) used for generating our training data. WMA was used with default parameters in our study. Evaluation metrics and results on testing datasets are as follows.

\subsubsection{SWM cluster identification rate}
For all compared methods, we quantified the SWM cluster identification rate (CIR) (Table \ref{tab_CIR}). The CIR is a metric that measures the success of white matter cluster identification in the absence of streamline labels \citep{Zhang2020-vm,Zhang2018-jx,Chen2021-dm,Chen2022-ow}. In our study, a cluster was considered to be successfully detected if there were at least 10 streamlines \citep{Zhang2018-jx,Zhang2020-vm}.  

Table \ref{tab_CIR} indicates that our proposed SupWMA method has the highest CIR on all six testing datasets, which demonstrates a good generalization of SupWMA to subjects across ages and health conditions. Although the training data only includes subjects from the healthy young adult population, CIRs for SupWMA are over 93\% on four testing datasets (HCP: healthy young adults; ABCD: healthy children; CNP: patients with psychiatric disorders; PPMI: healthy older adults and older adults with Parkinson’s disease).  CIRs of all methods decrease on two highly challenging datasets: BTP (patients with space-occupying brain tumors) and dHCP (neonates). CIRs are around 85\% for the BTP dataset due to patient-specific peritumoral effects. CIRs are below 70\% in the dHCP dataset, where the neonate brains do not yet have fully myelinated SWM \citep{Zhang2019-oq,Grotheer2022-rw}.

\begin{table*}[htbp]
\footnotesize
\caption{\textmd{Cluster spatial overlap (CSO) between compared methods and our method.}}
\centering
\begin{tabular}{|c|c|c|c|}
\hline
& WMA $\mid$ SupWMA & DeepWMA $\mid$ SupWMA  & DCNN+CL+ATT $\mid$ SupWMA \\
\hline
 HCP & 0.8290$\pm$0.1669 & 0.8740$\pm$0.1247 & 0.9232$\pm$0.0795\\
dHCP & 0.5131$\pm$0.3636 & 0.6401$\pm$0.3463 & 0.7601$\pm$0.2915\\
ABCD & 0.6834$\pm$0.2432 & 0.7762$\pm$0.2058 & 0.8320$\pm$0.1652\\
CNP & 0.7421$\pm$0.2306 & 0.8488$\pm$0.1731 & 0.9035$\pm$0.1267\\
PPMI & 0.8007$\pm$0.1992 & 0.8647$\pm$0.1592 & 0.9169$\pm$0.1142\\
BTP & 0.6984$\pm$0.2901 & 0.7605$\pm$0.2681 & 0.8415$\pm$0.2220\\
\hline
\end{tabular}
\label{tab_CSO}
\vspace{-2mm}  
\end{table*}

\begin{figure*}[!h] 
\centering
\includegraphics[width=1.0\textwidth]{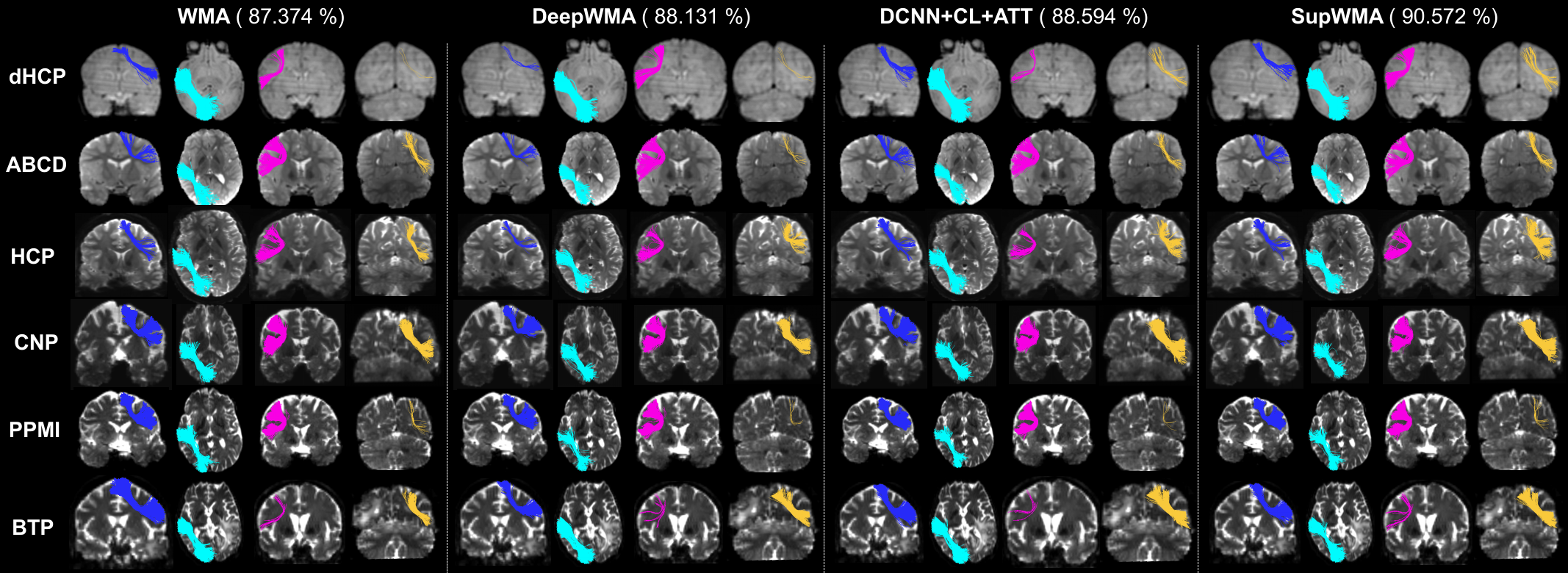}
\caption{\textmd{Visualization of example individual clusters. For each testing dataset, a representative subject (with mean CIR closest to the population mean) is selected for visualization. The average CIR (across the six representative subjects) is displayed for each method. Parcellated clusters and b0 images in the background were transferred into the atlas space for visualization.}}
\label{fig_visualization}
\end{figure*}

\subsubsection{Cluster distance to atlas}
For all compared methods, we quantified the SWM cluster distance to atlas (CDA) (Table \ref{tab_CDA}). The CDA aims to quantify the degree to which identified SWM clusters are geometrically similar to the corresponding atlas (training dataset) clusters, where a low CDA indicates high geometric similarity to the atlas. CDA leverages the well-known minimum average direct-flip (MDF) pairwise streamline distance, which is widely applied in white matter parcellation/clustering \citep{Garyfallidis2012-rs,Zhang2018-jx,Chen2021-dm}. For each streamline in the cluster of each testing subject, we calculated MDF distances between that streamline and all streamlines of the corresponding cluster of the atlas, and we recorded the minimum MDF distance. Then we computed the CDA as the average of these streamline-specific minimum MDF distances. For each method, Table \ref{tab_CDA} shows the average CDA across all identified SWM clusters of all testing subjects in all datasets.

As shown in Table \ref{tab_CDA}, the distance between identified SWM clusters and the corresponding clusters in the ORG atlas is low for all methods across all testing datasets (under 5.6 mm). This demonstrates that identified clusters are highly geometrically similar to clusters in the atlas in general. In addition, SupWMA and DCNN+CL+ATT are two best performers (lower CDA values than other methods) in all datasets, and SupWMA performs the best on three very challenging datasets: dHCP (neonates), ABCD (children) and BTP (tumor patients) datasets.

\subsubsection{Inter-subject parcellation variability}
For all compared methods, we quantified the inter-subject parcellation variability (ISPV) (Table \ref{tab_ISPV}). The ISPV is a metric for evaluating performance of tractography parcellation \citep{Roberts2017-hh,Zhang2017-ui,Zhang2018-jx} that evaluates if a cluster has a similar number of streamlines across subjects. The ISPV is defined as the coefficient of variation (standard deviation divided by the mean) of the number of streamlines per cluster. Therefore, the lower the ISPV, the higher the consistency of the number of streamlines of the corresponding cluster across subjects.

As shown in Table \ref{tab_ISPV}, SupWMA achieves highly consistent SWM parcellation results overall. It has lower ISPV than other compared methods on all testing datasets, especially on challenging dHCP (neonates), ABCD (children) and BTP (tumor patients) datasets, which are from individuals whose white matter anatomy is largely different from the training data population (healthy young adults). In addition, all methods have much higher ISPV values on dHCP  than other datasets potentially because of rapid brain development during the perinatal period \citep{Gilmore2018-mq}.

\subsubsection{Population cluster heatmap}

A population cluster heatmap displays the percentage of subjects that have streamlines present in each voxel. Here, population cluster heatmaps are generated to visualize and assess the cluster spatial overlap across subjects, as in \citep{Zhang2020-vm}. For computing heatmaps, clusters from different subjects and methods were registered (described in Section~\ref{dMRIDatasets}) into the space of a white matter atlas \citep{Zhang2018-jx}.

As shown in Fig. \ref{fig_heatmap}, identified clusters are visually comparable across the four SWM parcellation methods in general. Most clusters are identified robustly, as indicated by orange and red colors showing high cluster identification rates and high spatial overlap of clusters across subjects. However, more challenging clusters are less robustly identified by some methods, resulting in light orange and yellow colors that indicate lower identification of clusters that results in lower spatial overlap across subjects. We can observe that SupWMA obtains more consistent parcellation results than other methods, as indicated by visually higher~(more orange and red) voxel values in the heatmaps, especially on the dHCP dataset (neonates) and the clusters in the fourth column of each method. For this fourth column, it can be observed that WMA and DeepWMA almost fail to detect the clusters in the fourth column on the dHCP dataset.

\subsubsection{Cluster spatial overlap}
Cluster spatial overlap (CSO) is used to quantify if clusters from different methods are spatially comparable. In our experiments, CSO was quantified by comparing population cluster heatmaps using the weighted Dice (wDice) score, which has been used in many other tractography parcellation studies \citep{Cousineau2017-nf,Zhang2019-bl,Zhang2020-vm}. It extends the Dice score \citep{Taha2015-ij} to non-binary maps and assigns higher weighting to voxels with higher values.

As Table \ref{tab_CSO} shows, the CSO is generally high between methods across all datasets. This demonstrates that the results of SupWMA have good spatial overlap with other methods, and thus the results are comparable. SupWMA has higher CSO with the two deep learning methods (DeepWMA and DCNN+CL+ATT) than with the clustering-based method (WMA) on all testing datasets. A possible reason is that SupWMA and the two other deep learning methods were trained on the same dataset.

\subsubsection{Subject-specific cluster visualization}
We provide a visualization of the identified SWM clusters in an example individual subject for each dataset and across all SWM parcellation methods (Fig. \ref{fig_visualization}).

As shown in Fig. \ref{fig_visualization}, all methods have relatively good performance on clusters shown in blue, cyan and magenta. However, for yellow clusters, SupWMA identifies more streamlines that are visually plausible than other methods, especially on three datasets: dHCP (neonates), ABCD (children) and BTP (tumor patients). Also, it can be seen that SupWMA has better performance on the dHCP dataset (neonates) than other methods across the four selected clusters.

\subsubsection{Performance evaluation for clusters in peritumoral regions}
\begin{figure}[!h] 
\centering
\includegraphics[width=0.48\textwidth]{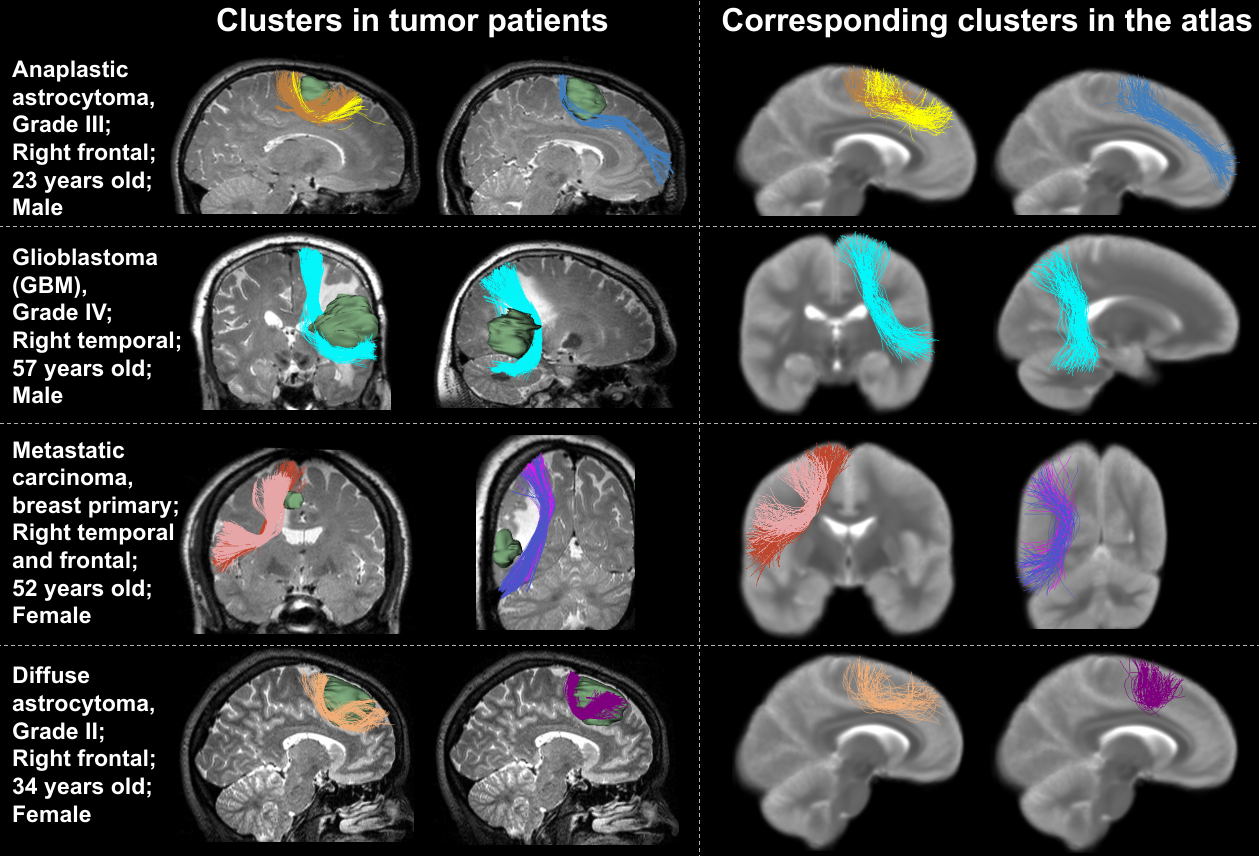}
\caption{\textmd{Visualization of SupWMA identified clusters located in the peritumoral region and corresponding clusters in the atlas. Four subjects with different diagnoses and tumor locations are selected, and clusters that have a large amount of overlap with edema and/or tumor are visualized. Brain tumors are shown in green, while each SWM cluster is displayed in a unique color.}}
\label{fig_tumor_vis}
\end{figure}

We provide a visualization of patient-specific SupWMA identified clusters in the peritumoral region, in selected patients from the BTP dataset (Fig.~\ref{fig_tumor_vis}). SupWMA can successfully identify peritumoral SWM clusters. Compared to clusters in the atlas, apparently atypical individual geometries are seen in identified SWM clusters, especially in patients with relatively larger tumors that may cause mass effect (e.g., the first and second rows of Fig.~\ref{fig_tumor_vis}).

To quantify the performance of SupWMA in the peritumoral region, we performed an additional experiment to calculate the identification rate of SWM clusters in the region of patient-specific tumor (n=39 patients) or edema (n=28 patients), as well as the  identification rate of SWM clusters within the normal appearing white matter (n=39 patients). First, expert segmentations of edema and tumor were performed and transformed to atlas space (using the transforms obtained in Section~\ref{dMRIDatasets}), and atlas clusters intersecting these segmentations were identified. This gave a list of expected peritumoral clusters (i.e., clusters likely to be proximal to tumor and/or edema) for each patient. Second, patient-specific identification rates in the peritumoral region were computed as the percentage of the expected peritumoral clusters that were identified in each patient (Table~\ref{tab_tumor_CIR}). The identification rate of all other clusters (in the normal-appearing white matter) was also computed for each patient (Table~\ref{tab_tumor_CIR}). Finally, for purposes of selecting peritumoral clusters for visualization, the patient-specific volume of overlap of each identified cluster with edema and tumor was computed.

\newcolumntype{P}[1]{>{\centering\arraybackslash}p{#1}} 
\newcolumntype{M}[1]{>{\centering\arraybackslash}m{#1}} 
\begin{table}[!h]
\footnotesize
\caption{\textmd{SWM cluster identification rates (CIRs) obtained using SupWMA for patient-specific clusters in the normal appearing white matter and the peritumoral region.}}
\centering
\begin{tabular}{|M{4.3cm}|M{3cm}|}
\hline
 &  Cluster identification rate \\
\hline
Clusters in normal-appearing ~~~~~~~~~~~~~~~~~~~white matter & \multicolumn{1}{c|}{92.98$\pm$6.76\%} \\
\hline
Patient-specific peritumoral clusters (in/near edema) & \multicolumn{1}{c|}{74.60$\pm$15.18\%} \\
\hline
Patient-specific peritumoral clusters (in/near tumor) & \multicolumn{1}{c|}{67.08$\pm$22.35\%} \\
\hline
\end{tabular}
\label{tab_tumor_CIR}
\end{table}

As shown in Table~\ref{tab_tumor_CIR}, SupWMA obtains a high SWM CIR (92.98\%) in the normal-appearing white matter. Although SWM connections may fail to be traced by tractography in peritumoral regions due to presence of mass lesions, SupWMA still identifies SWM clusters robustly with SWM CIR of 74.6\% for clusters in/near peritumoral edema, and 67.08\% for clusters nearest the tumor.

\subsubsection{Computation time}
Computation time was tested on a Linux workstation with CPU~(AMD Ryzen 5 3600) and GPU~(NVIDIA RTX 2080 Ti) using a randomly selected subject (0.44 million streamlines). Table \ref{tab_computation_time} shows the computation time across all compared SWM parcellation methods. Our method has the shortest computation time using CPU only and using CPU plus GPU. Also, benefiting from the efficient network structure, our method has the smallest computation time increase when changing from CPU plus GPU mode to CPU only mode.

\begin{table}[!h]
\footnotesize
\caption{\textmd{Comparisons of computation time across methods.}}
\centering
\begin{tabular}{|c|c|c|}
\hline
Methods & CPU only & CPU + GPU  \\
\hline
WMA & 101min & |  \\
DeepWMA & 4min17s & 2min22s \\
DCNN+CL+ATT & 4min51s & 1min51s \\
SupWMA & \textbf{1min52s} & \textbf{1min16s} \\
\hline
\end{tabular}
\label{tab_computation_time}
\end{table}

\section{Discussion and conclusion}
\label{conclusion}
In this study, we proposed SupWMA, a novel deep learning framework for SWM parcellation, with successful application on datasets across image acquisition protocols, ages and health conditions. SupWMA enables efficient and consistent SWM parcellation.

SWM parcellation is a challenging problem due to the small diameter of SWM fasciculi, their short fiber length, their high variability across subjects, and their unique position near the cortex. As the first deep learning method for SWM parcellation, SupWMA can efficiently generate consistent SWM parcellation results to benefit downstream neuroscientific studies. These studies rely on diffusion features (FA, MD, etc.) of tracts from SWM parcellation \citep{Wu2014-ai,Reginold2016-zq,DAlbis2018-ic,Ji2018-hq,Hatton2014-sf}.

Unlike other deep learning methods \citep{Xu2019-aa,Liu2019-jj,Ngattai_Lam2018-ne} for white matter parcellation, SupWMA encodes geometric features of points on streamlines using a point-cloud-based network, which allows streamlines with equivalent forward and reverse orders to have the same representation. The two-stage framework design improves performance by decomposing the complicated SWM parcellation into two easier subtasks and enabling the SCL training to focus on the SWM. The incorporation of SCL enhances the SWM parcellation performance by helping the encoder to output discriminative representations of streamlines. Also, the removal of transformation nets preserves significant spatial features and reduces the computation time of training and testing processes.  Compared to existing SOTA methods, SupWMA is the top performer on the training dataset and on six independently acquired testing datasets.

There are also several limitations and directions for future work. First, SupWMA has leveraged an anatomically curated atlas that includes SWM clusters discovered in a data-driven way in a population of healthy adults. While the employed atlas provides more comprehensive coverage of the SWM compared to other atlases \citep{Guevara2017-gz,Roman2017-uy}, it should be noted that it focuses on the long- and medium-range SWM connections (over 40~mm in length). Recent research has shown the potential of high-resolution dMRI scans to enable tractography of shorter, highly curved u-fibers of the SWM \citep{Song2014-un,Ramos-Llorden2020-vt}. Benefitting from that, future work can utilize new training data (SWM atlases) to identify shorter connections, which are nearer the cortex and hence more variable across subjects \citep{Roman2021-rd,Roman2022-pr}.~Second, the proposed method is designed to enable studies of brain symmetry and asymmetry, as the definition of the SWM clusters is performed in a bilateral fashion. However, while corresponding SWM clusters in the left and right hemispheres have highly similar geometry and generally intersect corresponding cortical regions (Supplementary Material 2), leveraging additional anatomical information can potentially improve clustering results for the creation of new SWM atlases \citep{Chen2021-dm}. Third, a potential limitation of our method is that only one atlas dataset was used for training. This strategy of training models on one dataset (e.g., HCP) and applying trained models on other datasets has been widely used in deep white matter parcellation/segmentation \citep{Wasserthal2018-if,Wasserthal2019-ha,Zhang2020-vm,Lu2022-ym} and dMRI registration \citep{Zhang2022-vc}. Furthermore, related work in SWM parcellation using non-deep methods traditionally uses one atlas dataset to define the parcellation \citep{Guevara2017-gz,Roman2021-rd,Roman2022-pr}. In this work, we have used an atlas that provides comprehensive coverage of the SWM and a fine parcellation scale \citep{Zhang2018-jx}. Fourth, another limitation of our method is that it may fail to identify SWM connections that are not traced by tractography (due to issues such as the presence of mass lesions, ongoing neurodevelopment, or differences in scan protocol). However, SupWMA identified clusters robustly for all datasets overall. In the challenging BTP dataset, cluster identification was generally robust even in the peritumoral region (Fig.~\ref{fig_tumor_vis} and Table~\ref{tab_tumor_CIR}). In the worst-case cluster performance in the challenging dHCP dataset, where neurodevelopment is ongoing and scanning is challenging, the cluster with the minimum identification rate was still found in 10\% of subjects (Supplementary Material 3). Fifth, performing streamline tractography near the cortex is challenging due to the crossing of long-range fibers through superficial fiber systems \citep{Reveley2015-ki} and to the bias where tractography algorithms preferentially terminate in the crowns of gyri~\citep{Schilling2018-el,Van_Essen2014-mc}. While addressing these fundamental anatomical challenges in performing tractography is beyond the scope of the current work, we note that future tractography methods that increase cortical coverage of streamlines \citep{Bastiani2017-pl,St-Onge2018-rw,Wu2020-hn} will affect streamline endpoints. As the endpoints were shown to be important for SWM parcellation, this may improve depiction and discrimination of small bundles near the cortex. Sixth,~more advanced point-cloud-based network structures \citep{Qi2017-dw,Wang2018-xa} can be applied to improve the performance.

\section*{Code and data availability}
The code, the trained model, and the training dataset will be made available at: \href{https://supwma.github.io}{https://supwma.github.io}.

\section*{Acknowledgments}
We thank Erickson Torio, Shun Gong, Walid I. Essayed, Prashin Unadkat and Laura Rigolo for their help acquiring and segmenting the tumor patient dataset. We acknowledge the following NIH grants: P41EB015902, R01MH074794, R01MH125860, R01NS125781, R01NS125307, and R01MH119222. F.Z. also acknowledges a BWH Radiology Research Pilot Grant Award.





\bibliographystyle{model2-names.bst}\biboptions{authoryear}
\bibliography{refs}

\begin{thebibliography}{93}
\expandafter\ifx\csname natexlab\endcsname\relax\def\natexlab#1{#1}\fi
\providecommand{\url}[1]{\texttt{#1}}
\providecommand{\href}[2]{#2}
\providecommand{\path}[1]{#1}
\providecommand{\DOIprefix}{doi:}
\providecommand{\ArXivprefix}{arXiv:}
\providecommand{\URLprefix}{URL: }
\providecommand{\Pubmedprefix}{pmid:}
\providecommand{\doi}[1]{\href{http://dx.doi.org/#1}{\path{#1}}}
\providecommand{\Pubmed}[1]{\href{pmid:#1}{\path{#1}}}
\providecommand{\bibinfo}[2]{#2}
\ifx\xfnm\relax \def\xfnm[#1]{\unskip,\space#1}\fi
\bibitem[{Al-masni et~al.(2020)Al-masni, Kim, Kim, Noh and
  Kim}]{Al-masni2020-oy}
\bibinfo{author}{Al-masni, M.A.}, \bibinfo{author}{Kim, W.R.},
  \bibinfo{author}{Kim, E.Y.}, \bibinfo{author}{Noh, Y.}, \bibinfo{author}{Kim,
  D.H.}, \bibinfo{year}{2020}.
\newblock \bibinfo{title}{Automated detection of cerebral microbleeds in {MR}
  images: A two-stage deep learning approach}.
\newblock \bibinfo{journal}{NeuroImage: Clinical} \bibinfo{volume}{28},
  \bibinfo{pages}{102464}.
\bibitem[{Alexander et~al.(2001)Alexander, Hasan, Lazar, Tsuruda and
  Parker}]{Alexander2001-ai}
\bibinfo{author}{Alexander, A.L.}, \bibinfo{author}{Hasan, K.M.},
  \bibinfo{author}{Lazar, M.}, \bibinfo{author}{Tsuruda, J.S.},
  \bibinfo{author}{Parker, D.L.}, \bibinfo{year}{2001}.
\newblock \bibinfo{title}{Analysis of partial volume effects in
  diffusion-tensor {MRI}}.
\newblock \bibinfo{journal}{Magnetic Resonance in Medicine}
  \bibinfo{volume}{45}, \bibinfo{pages}{770--780}.
\bibitem[{Astolfi et~al.(2020a)Astolfi, De~Benedictis, Sarubbo, Bert{\'o},
  Olivetti, Sona and Avesani}]{Astolfi2020-zk}
\bibinfo{author}{Astolfi, P.}, \bibinfo{author}{De~Benedictis, A.},
  \bibinfo{author}{Sarubbo, S.}, \bibinfo{author}{Bert{\'o}, G.},
  \bibinfo{author}{Olivetti, E.}, \bibinfo{author}{Sona, D.},
  \bibinfo{author}{Avesani, P.}, \bibinfo{year}{2020}a.
\newblock \bibinfo{title}{A {Stem-Based} dissection of inferior
  {Fronto-Occipital} fasciculus with a deep learning model}, in:
  \bibinfo{booktitle}{2020 {IEEE} 17th International Symposium on Biomedical
  Imaging ({ISBI})}, pp. \bibinfo{pages}{267--270}.
\bibitem[{Astolfi et~al.(2020b)Astolfi, Verhagen, Petit, Olivetti, Masci,
  Boscaini and Avesani}]{Astolfi2020-jf}
\bibinfo{author}{Astolfi, P.}, \bibinfo{author}{Verhagen, R.},
  \bibinfo{author}{Petit, L.}, \bibinfo{author}{Olivetti, E.},
  \bibinfo{author}{Masci, J.}, \bibinfo{author}{Boscaini, D.},
  \bibinfo{author}{Avesani, P.}, \bibinfo{year}{2020}b.
\newblock \bibinfo{title}{Tractogram filtering of anatomically non-plausible
  fibers with geometric deep learning}, in: \bibinfo{booktitle}{Medical Image
  Computing and Computer Assisted Intervention (MICCAI)}, pp.
  \bibinfo{pages}{291--301}.
\bibitem[{Basser et~al.(2000)Basser, Pajevic, Pierpaoli, Duda and
  Aldroubi}]{Basser2000-tu}
\bibinfo{author}{Basser, P.J.}, \bibinfo{author}{Pajevic, S.},
  \bibinfo{author}{Pierpaoli, C.}, \bibinfo{author}{Duda, J.},
  \bibinfo{author}{Aldroubi, A.}, \bibinfo{year}{2000}.
\newblock \bibinfo{title}{In vivo fiber tractography using {DT-MRI} data}.
\newblock \bibinfo{journal}{Magn. Reson. Med.} \bibinfo{volume}{44},
  \bibinfo{pages}{625--632}.
\bibitem[{Bastiani et~al.(2017)Bastiani, Cottaar, Dikranian, Ghosh, Zhang,
  Alexander, Behrens, Jbabdi and Sotiropoulos}]{Bastiani2017-pl}
\bibinfo{author}{Bastiani, M.}, \bibinfo{author}{Cottaar, M.},
  \bibinfo{author}{Dikranian, K.}, \bibinfo{author}{Ghosh, A.},
  \bibinfo{author}{Zhang, H.}, \bibinfo{author}{Alexander, D.C.},
  \bibinfo{author}{Behrens, T.E.}, \bibinfo{author}{Jbabdi, S.},
  \bibinfo{author}{Sotiropoulos, S.N.}, \bibinfo{year}{2017}.
\newblock \bibinfo{title}{Improved tractography using asymmetric fibre
  orientation distributions}.
\newblock \bibinfo{journal}{Neuroimage} \bibinfo{volume}{158},
  \bibinfo{pages}{205--218}.
\bibitem[{Charles et~al.(2017)Charles, Su, Kaichun and Guibas}]{Charles2017-nv}
\bibinfo{author}{Charles, R.Q.}, \bibinfo{author}{Su, H.},
  \bibinfo{author}{Kaichun, M.}, \bibinfo{author}{Guibas, L.J.},
  \bibinfo{year}{2017}.
\newblock \bibinfo{title}{{PointNet}: Deep learning on point sets for {3D}
  classification and segmentation}, in: \bibinfo{booktitle}{2017 {IEEE}
  Conference on Computer Vision and Pattern Recognition ({CVPR})}, pp.
  \bibinfo{pages}{77--85}.
\bibitem[{Chen et~al.(2020)Chen, Kornblith, Norouzi and Hinton}]{Chen2020-pz}
\bibinfo{author}{Chen, T.}, \bibinfo{author}{Kornblith, S.},
  \bibinfo{author}{Norouzi, M.}, \bibinfo{author}{Hinton, G.},
  \bibinfo{year}{2020}.
\newblock \bibinfo{title}{A simple framework for contrastive learning of visual
  representations}, in: \bibinfo{editor}{Iii, H.D.}, \bibinfo{editor}{Singh,
  A.} (Eds.), \bibinfo{booktitle}{Proceedings of the 37th International
  Conference on Machine Learning (ICML)}, pp. \bibinfo{pages}{1597--1607}.
\bibitem[{Chen et~al.(2021)Chen, Zhang, Song, Makris, Rathi, Cai, Zhang and
  O'Donnell}]{Chen2021-dm}
\bibinfo{author}{Chen, Y.}, \bibinfo{author}{Zhang, C.}, \bibinfo{author}{Song,
  Y.}, \bibinfo{author}{Makris, N.}, \bibinfo{author}{Rathi, Y.},
  \bibinfo{author}{Cai, W.}, \bibinfo{author}{Zhang, F.},
  \bibinfo{author}{O'Donnell, L.J.}, \bibinfo{year}{2021}.
\newblock \bibinfo{title}{Deep fiber clustering: Anatomically informed
  unsupervised deep learning for fast and effective white matter parcellation},
  in: \bibinfo{booktitle}{Medical Image Computing and Computer Assisted
  Intervention (MICCAI)}, pp. \bibinfo{pages}{497--507}.
\bibitem[{Chen et~al.(2022)Chen, Zhang, Xue, Song, Makris, Rathi, Cai, Zhang
  and O'Donnell}]{Chen2022-ow}
\bibinfo{author}{Chen, Y.}, \bibinfo{author}{Zhang, C.}, \bibinfo{author}{Xue,
  T.}, \bibinfo{author}{Song, Y.}, \bibinfo{author}{Makris, N.},
  \bibinfo{author}{Rathi, Y.}, \bibinfo{author}{Cai, W.},
  \bibinfo{author}{Zhang, F.}, \bibinfo{author}{O'Donnell, L.J.},
  \bibinfo{year}{2022}.
\newblock \bibinfo{title}{{DFC}: Anatomically informed fiber clustering with
  self-supervised deep learning for fast and effective tractography
  parcellation} \href{http://arxiv.org/abs/2205.00627}{\tt arXiv:2205.00627}.
\bibitem[{Chen et~al.(2015)Chen, Tie, Olubiyi, Rigolo, Mehrtash, Norton,
  Pasternak, Rathi, Golby and O'Donnell}]{Chen2015-fy}
\bibinfo{author}{Chen, Z.}, \bibinfo{author}{Tie, Y.},
  \bibinfo{author}{Olubiyi, O.}, \bibinfo{author}{Rigolo, L.},
  \bibinfo{author}{Mehrtash, A.}, \bibinfo{author}{Norton, I.},
  \bibinfo{author}{Pasternak, O.}, \bibinfo{author}{Rathi, Y.},
  \bibinfo{author}{Golby, A.J.}, \bibinfo{author}{O'Donnell, L.J.},
  \bibinfo{year}{2015}.
\newblock \bibinfo{title}{Reconstruction of the arcuate fasciculus for surgical
  planning in the setting of peritumoral edema using two-tensor unscented
  kalman filter tractography}.
\newblock \bibinfo{journal}{Neuroimage Clin} \bibinfo{volume}{7},
  \bibinfo{pages}{815--822}.
\bibitem[{Cousineau et~al.(2017)Cousineau, Jodoin, Morency, Rozanski,
  Grand'Maison, Bedell and Descoteaux}]{Cousineau2017-nf}
\bibinfo{author}{Cousineau, M.}, \bibinfo{author}{Jodoin, P.M.},
  \bibinfo{author}{Morency, F.C.}, \bibinfo{author}{Rozanski, V.},
  \bibinfo{author}{Grand'Maison, M.}, \bibinfo{author}{Bedell, B.J.},
  \bibinfo{author}{Descoteaux, M.}, \bibinfo{year}{2017}.
\newblock \bibinfo{title}{A test-retest study on parkinson's {PPMI} dataset
  yields statistically significant white matter fascicles}.
\newblock \bibinfo{journal}{Neuroimage Clin} \bibinfo{volume}{16},
  \bibinfo{pages}{222--233}.
\bibitem[{d'Albis et~al.(2018)d'Albis, Guevara, Guevara, Laidi, Boisgontier,
  Sarrazin, Duclap, Delorme, Bolognani, Czech, Bouquet, Ly-Le~Moal, Holiga,
  Amestoy, Scheid, Gaman, Leboyer, Poupon, Mangin and Houenou}]{DAlbis2018-ic}
\bibinfo{author}{d'Albis, M.A.}, \bibinfo{author}{Guevara, P.},
  \bibinfo{author}{Guevara, M.}, \bibinfo{author}{Laidi, C.},
  \bibinfo{author}{Boisgontier, J.}, \bibinfo{author}{Sarrazin, S.},
  \bibinfo{author}{Duclap, D.}, \bibinfo{author}{Delorme, R.},
  \bibinfo{author}{Bolognani, F.}, \bibinfo{author}{Czech, C.},
  \bibinfo{author}{Bouquet, C.}, \bibinfo{author}{Ly-Le~Moal, M.},
  \bibinfo{author}{Holiga, S.}, \bibinfo{author}{Amestoy, A.},
  \bibinfo{author}{Scheid, I.}, \bibinfo{author}{Gaman, A.},
  \bibinfo{author}{Leboyer, M.}, \bibinfo{author}{Poupon, C.},
  \bibinfo{author}{Mangin, J.F.}, \bibinfo{author}{Houenou, J.},
  \bibinfo{year}{2018}.
\newblock \bibinfo{title}{Local structural connectivity is associated with
  social cognition in autism spectrum disorder}.
\newblock \bibinfo{journal}{Brain} \bibinfo{volume}{141},
  \bibinfo{pages}{3472--3481}.
\bibitem[{Edwards et~al.(2022)Edwards, Rueckert, Smith, Abo~Seada, Alansary,
  Almalbis, Allsop, Andersson, Arichi, Arulkumaran, Bastiani, Batalle, Baxter,
  Bozek, Braithwaite, Brandon, Carney, Chew, Christiaens, Chung, Colford,
  Cordero-Grande, Counsell, Cullen, Cupitt, Curtis, Davidson, Deprez, Dillon,
  Dimitrakopoulou, Dimitrova, Duff, Falconer, Farahibozorg, Fitzgibbon, Gao,
  Gaspar, Harper, Harrison, Hughes, Hutter, Jenkinson, Jbabdi, Jones, Karolis,
  Kyriakopoulou, Lenz, Makropoulos, Malik, Mason, Mortari, Nosarti, Nunes,
  O'Keeffe, O'Muircheartaigh, Patel, Passerat-Palmbach, Pietsch, Price,
  Robinson, Rutherford, Schuh, Sotiropoulos, Steinweg, Teixeira, Tenev,
  Tournier, Tusor, Uus, Vecchiato, Williams, Wright, Wurie and
  Hajnal}]{Edwards2022-vr}
\bibinfo{author}{Edwards, A.D.}, \bibinfo{author}{Rueckert, D.},
  \bibinfo{author}{Smith, S.M.}, \bibinfo{author}{Abo~Seada, S.},
  \bibinfo{author}{Alansary, A.}, \bibinfo{author}{Almalbis, J.},
  \bibinfo{author}{Allsop, J.}, \bibinfo{author}{Andersson, J.},
  \bibinfo{author}{Arichi, T.}, \bibinfo{author}{Arulkumaran, S.},
  \bibinfo{author}{Bastiani, M.}, \bibinfo{author}{Batalle, D.},
  \bibinfo{author}{Baxter, L.}, \bibinfo{author}{Bozek, J.},
  \bibinfo{author}{Braithwaite, E.}, \bibinfo{author}{Brandon, J.},
  \bibinfo{author}{Carney, O.}, \bibinfo{author}{Chew, A.},
  \bibinfo{author}{Christiaens, D.}, \bibinfo{author}{Chung, R.},
  \bibinfo{author}{Colford, K.}, \bibinfo{author}{Cordero-Grande, L.},
  \bibinfo{author}{Counsell, S.J.}, \bibinfo{author}{Cullen, H.},
  \bibinfo{author}{Cupitt, J.}, \bibinfo{author}{Curtis, C.},
  \bibinfo{author}{Davidson, A.}, \bibinfo{author}{Deprez, M.},
  \bibinfo{author}{Dillon, L.}, \bibinfo{author}{Dimitrakopoulou, K.},
  \bibinfo{author}{Dimitrova, R.}, \bibinfo{author}{Duff, E.},
  \bibinfo{author}{Falconer, S.}, \bibinfo{author}{Farahibozorg, S.R.},
  \bibinfo{author}{Fitzgibbon, S.P.}, \bibinfo{author}{Gao, J.},
  \bibinfo{author}{Gaspar, A.}, \bibinfo{author}{Harper, N.},
  \bibinfo{author}{Harrison, S.J.}, \bibinfo{author}{Hughes, E.J.},
  \bibinfo{author}{Hutter, J.}, \bibinfo{author}{Jenkinson, M.},
  \bibinfo{author}{Jbabdi, S.}, \bibinfo{author}{Jones, E.},
  \bibinfo{author}{Karolis, V.}, \bibinfo{author}{Kyriakopoulou, V.},
  \bibinfo{author}{Lenz, G.}, \bibinfo{author}{Makropoulos, A.},
  \bibinfo{author}{Malik, S.}, \bibinfo{author}{Mason, L.},
  \bibinfo{author}{Mortari, F.}, \bibinfo{author}{Nosarti, C.},
  \bibinfo{author}{Nunes, R.G.}, \bibinfo{author}{O'Keeffe, C.},
  \bibinfo{author}{O'Muircheartaigh, J.}, \bibinfo{author}{Patel, H.},
  \bibinfo{author}{Passerat-Palmbach, J.}, \bibinfo{author}{Pietsch, M.},
  \bibinfo{author}{Price, A.N.}, \bibinfo{author}{Robinson, E.C.},
  \bibinfo{author}{Rutherford, M.A.}, \bibinfo{author}{Schuh, A.},
  \bibinfo{author}{Sotiropoulos, S.}, \bibinfo{author}{Steinweg, J.},
  \bibinfo{author}{Teixeira, R.P.A.G.}, \bibinfo{author}{Tenev, T.},
  \bibinfo{author}{Tournier, J.D.}, \bibinfo{author}{Tusor, N.},
  \bibinfo{author}{Uus, A.}, \bibinfo{author}{Vecchiato, K.},
  \bibinfo{author}{Williams, L.Z.J.}, \bibinfo{author}{Wright, R.},
  \bibinfo{author}{Wurie, J.}, \bibinfo{author}{Hajnal, J.V.},
  \bibinfo{year}{2022}.
\newblock \bibinfo{title}{The developing human connectome project neonatal data
  release}.
\newblock \bibinfo{journal}{Front. Neurosci.} \bibinfo{volume}{16},
  \bibinfo{pages}{886772}.
\bibitem[{Fedorov et~al.(2012)Fedorov, Beichel, Kalpathy-Cramer, Finet,
  Fillion-Robin, Pujol, Bauer, Jennings, Fennessy, Sonka, Buatti, Aylward,
  Miller, Pieper and Kikinis}]{Fedorov2012-ma}
\bibinfo{author}{Fedorov, A.}, \bibinfo{author}{Beichel, R.},
  \bibinfo{author}{Kalpathy-Cramer, J.}, \bibinfo{author}{Finet, J.},
  \bibinfo{author}{Fillion-Robin, J.C.}, \bibinfo{author}{Pujol, S.},
  \bibinfo{author}{Bauer, C.}, \bibinfo{author}{Jennings, D.},
  \bibinfo{author}{Fennessy, F.}, \bibinfo{author}{Sonka, M.},
  \bibinfo{author}{Buatti, J.}, \bibinfo{author}{Aylward, S.},
  \bibinfo{author}{Miller, J.V.}, \bibinfo{author}{Pieper, S.},
  \bibinfo{author}{Kikinis, R.}, \bibinfo{year}{2012}.
\newblock \bibinfo{title}{{3D} slicer as an image computing platform for the
  quantitative imaging network}.
\newblock \bibinfo{journal}{Magn. Reson. Imaging} \bibinfo{volume}{30},
  \bibinfo{pages}{1323--1341}.
\bibitem[{Garyfallidis et~al.(2012)Garyfallidis, Brett, Correia, Williams and
  Nimmo-Smith}]{Garyfallidis2012-rs}
\bibinfo{author}{Garyfallidis, E.}, \bibinfo{author}{Brett, M.},
  \bibinfo{author}{Correia, M.M.}, \bibinfo{author}{Williams, G.B.},
  \bibinfo{author}{Nimmo-Smith, I.}, \bibinfo{year}{2012}.
\newblock \bibinfo{title}{{QuickBundles}, a method for tractography
  simplification}.
\newblock \bibinfo{journal}{Front. Neurosci.} \bibinfo{volume}{6},
  \bibinfo{pages}{175}.
\bibitem[{Garyfallidis et~al.(2018)Garyfallidis, C{\^o}t{\'e}, Rheault, Sidhu,
  Hau, Petit, Fortin, Cunanne and Descoteaux}]{Garyfallidis2018-hn}
\bibinfo{author}{Garyfallidis, E.}, \bibinfo{author}{C{\^o}t{\'e}, M.A.},
  \bibinfo{author}{Rheault, F.}, \bibinfo{author}{Sidhu, J.},
  \bibinfo{author}{Hau, J.}, \bibinfo{author}{Petit, L.},
  \bibinfo{author}{Fortin, D.}, \bibinfo{author}{Cunanne, S.},
  \bibinfo{author}{Descoteaux, M.}, \bibinfo{year}{2018}.
\newblock \bibinfo{title}{Recognition of white matter bundles using local and
  global streamline-based registration and clustering}.
\newblock \bibinfo{journal}{Neuroimage} \bibinfo{volume}{170},
  \bibinfo{pages}{283--295}.
\bibitem[{Gilmore et~al.(2018)Gilmore, Knickmeyer and Gao}]{Gilmore2018-mq}
\bibinfo{author}{Gilmore, J.H.}, \bibinfo{author}{Knickmeyer, R.C.},
  \bibinfo{author}{Gao, W.}, \bibinfo{year}{2018}.
\newblock \bibinfo{title}{Imaging structural and functional brain development
  in early childhood}.
\newblock \bibinfo{journal}{Nat. Rev. Neurosci.} \bibinfo{volume}{19},
  \bibinfo{pages}{123--137}.
\bibitem[{Grotheer et~al.(2022)Grotheer, Rosenke, Wu, Kular, Querdasi, Natu,
  Yeatman and Grill-Spector}]{Grotheer2022-rw}
\bibinfo{author}{Grotheer, M.}, \bibinfo{author}{Rosenke, M.},
  \bibinfo{author}{Wu, H.}, \bibinfo{author}{Kular, H.},
  \bibinfo{author}{Querdasi, F.R.}, \bibinfo{author}{Natu, V.S.},
  \bibinfo{author}{Yeatman, J.D.}, \bibinfo{author}{Grill-Spector, K.},
  \bibinfo{year}{2022}.
\newblock \bibinfo{title}{White matter myelination during early infancy is
  linked to spatial gradients and myelin content at birth}.
\newblock \bibinfo{journal}{Nat. Commun.} \bibinfo{volume}{13},
  \bibinfo{pages}{997}.
\bibitem[{Guevara et~al.(2020)Guevara, Guevara, Rom{\'a}n and
  Mangin}]{Guevara2020-da}
\bibinfo{author}{Guevara, M.}, \bibinfo{author}{Guevara, P.},
  \bibinfo{author}{Rom{\'a}n, C.}, \bibinfo{author}{Mangin, J.F.},
  \bibinfo{year}{2020}.
\newblock \bibinfo{title}{Superficial white matter: A review on the {dMRI}
  analysis methods and applications}.
\newblock \bibinfo{journal}{Neuroimage} \bibinfo{volume}{212},
  \bibinfo{pages}{116673}.
\bibitem[{Guevara et~al.(2017)Guevara, Rom{\'a}n, Houenou, Duclap, Poupon,
  Mangin and Guevara}]{Guevara2017-gz}
\bibinfo{author}{Guevara, M.}, \bibinfo{author}{Rom{\'a}n, C.},
  \bibinfo{author}{Houenou, J.}, \bibinfo{author}{Duclap, D.},
  \bibinfo{author}{Poupon, C.}, \bibinfo{author}{Mangin, J.F.},
  \bibinfo{author}{Guevara, P.}, \bibinfo{year}{2017}.
\newblock \bibinfo{title}{Reproducibility of superficial white matter tracts
  using diffusion-weighted imaging tractography}.
\newblock \bibinfo{journal}{Neuroimage} \bibinfo{volume}{147},
  \bibinfo{pages}{703--725}.
\bibitem[{Guevara et~al.(2012)Guevara, Duclap, Poupon, Marrakchi-Kacem,
  Fillard, Le~Bihan, Leboyer, Houenou and Mangin}]{Guevara2012-ex}
\bibinfo{author}{Guevara, P.}, \bibinfo{author}{Duclap, D.},
  \bibinfo{author}{Poupon, C.}, \bibinfo{author}{Marrakchi-Kacem, L.},
  \bibinfo{author}{Fillard, P.}, \bibinfo{author}{Le~Bihan, D.},
  \bibinfo{author}{Leboyer, M.}, \bibinfo{author}{Houenou, J.},
  \bibinfo{author}{Mangin, J.F.}, \bibinfo{year}{2012}.
\newblock \bibinfo{title}{Automatic fiber bundle segmentation in massive
  tractography datasets using a multi-subject bundle atlas}.
\newblock \bibinfo{journal}{Neuroimage} \bibinfo{volume}{61},
  \bibinfo{pages}{1083--1099}.
\bibitem[{Gunel et~al.(2021)Gunel, Du, Conneau and Stoyanov}]{Gunel2020-zq}
\bibinfo{author}{Gunel, B.}, \bibinfo{author}{Du, J.},
  \bibinfo{author}{Conneau, A.}, \bibinfo{author}{Stoyanov, V.},
  \bibinfo{year}{2021}.
\newblock \bibinfo{title}{Supervised contrastive learning for pre-trained
  language model fine-tuning}, in: \bibinfo{booktitle}{International Conference
  on Learning Representations (ICLR)}.
\bibitem[{Guo et~al.(2020)Guo, Wang, Hu, Liu, Liu and Bennamoun}]{Guo2020-ql}
\bibinfo{author}{Guo, Y.}, \bibinfo{author}{Wang, H.}, \bibinfo{author}{Hu,
  Q.}, \bibinfo{author}{Liu, H.}, \bibinfo{author}{Liu, L.},
  \bibinfo{author}{Bennamoun, M.}, \bibinfo{year}{2020}.
\newblock \bibinfo{title}{Deep learning for {3D} point clouds: A survey}.
\newblock \bibinfo{journal}{IEEE Trans. Pattern Anal. Mach. Intell.}
  \bibinfo{volume}{PP}.
\bibitem[{Han et~al.(2021)Han, Huang, Yang and Han}]{Han2021-ke}
\bibinfo{author}{Han, T.}, \bibinfo{author}{Huang, H.}, \bibinfo{author}{Yang,
  Z.}, \bibinfo{author}{Han, W.}, \bibinfo{year}{2021}.
\newblock \bibinfo{title}{Supervised contrastive learning for accented speech
  recognition} \href{http://arxiv.org/abs/2107.00921}{\tt arXiv:2107.00921}.
\bibitem[{Hatton et~al.(2014)Hatton, Lagopoulos, Hermens, Hickie, Scott and
  Bennett}]{Hatton2014-sf}
\bibinfo{author}{Hatton, S.N.}, \bibinfo{author}{Lagopoulos, J.},
  \bibinfo{author}{Hermens, D.F.}, \bibinfo{author}{Hickie, I.B.},
  \bibinfo{author}{Scott, E.}, \bibinfo{author}{Bennett, M.R.},
  \bibinfo{year}{2014}.
\newblock \bibinfo{title}{Short association fibres of the
  insula-temporoparietal junction in early psychosis: a diffusion tensor
  imaging study}.
\newblock \bibinfo{journal}{PLoS One} \bibinfo{volume}{9},
  \bibinfo{pages}{e112842}.
\bibitem[{He et~al.(2016)He, Zhang, Ren and Sun}]{He2015-nc}
\bibinfo{author}{He, K.}, \bibinfo{author}{Zhang, X.}, \bibinfo{author}{Ren,
  S.}, \bibinfo{author}{Sun, J.}, \bibinfo{year}{2016}.
\newblock \bibinfo{title}{Deep residual learning for image recognition}, in:
  \bibinfo{booktitle}{Proceedings of the IEEE conference on computer vision and
  pattern recognition (CVPR)}, pp. \bibinfo{pages}{770--778}.
\bibitem[{Huang et~al.(2021)Huang, Ko, Lilian~Tang, Liu and Wu}]{Huang2021-fl}
\bibinfo{author}{Huang, Q.}, \bibinfo{author}{Ko, T.},
  \bibinfo{author}{Lilian~Tang, H.}, \bibinfo{author}{Liu, X.},
  \bibinfo{author}{Wu, B.}, \bibinfo{year}{2021}.
\newblock \bibinfo{title}{Token-level supervised contrastive learning for
  punctuation restoration}, in: \bibinfo{booktitle}{Annual Conference of the
  International Speech Communication Association}.
\bibitem[{Jain et~al.(2020)Jain, Gajbhiye, Tripathy and Acharya}]{Jain2020-cb}
\bibinfo{author}{Jain, P.}, \bibinfo{author}{Gajbhiye, P.},
  \bibinfo{author}{Tripathy, R.K.}, \bibinfo{author}{Acharya, U.R.},
  \bibinfo{year}{2020}.
\newblock \bibinfo{title}{A two-stage deep {CNN} architecture for the
  classification of low-risk and high-risk hypertension classes using
  multi-lead {ECG} signals}.
\newblock \bibinfo{journal}{Informatics in Medicine Unlocked}
  \bibinfo{volume}{21}, \bibinfo{pages}{100479}.
\bibitem[{Ji et~al.(2019)Ji, Guevara, Guevara, Grigis, Labra, Sarrazin,
  Hamdani, Bellivier, Delavest, Leboyer, Tamouza, Poupon, Mangin and
  Houenou}]{Ji2019-ie}
\bibinfo{author}{Ji, E.}, \bibinfo{author}{Guevara, P.},
  \bibinfo{author}{Guevara, M.}, \bibinfo{author}{Grigis, A.},
  \bibinfo{author}{Labra, N.}, \bibinfo{author}{Sarrazin, S.},
  \bibinfo{author}{Hamdani, N.}, \bibinfo{author}{Bellivier, F.},
  \bibinfo{author}{Delavest, M.}, \bibinfo{author}{Leboyer, M.},
  \bibinfo{author}{Tamouza, R.}, \bibinfo{author}{Poupon, C.},
  \bibinfo{author}{Mangin, J.F.}, \bibinfo{author}{Houenou, J.},
  \bibinfo{year}{2019}.
\newblock \bibinfo{title}{Increased and decreased superficial white matter
  structural connectivity in schizophrenia and bipolar disorder}.
\newblock \bibinfo{journal}{Schizophr. Bull.} \bibinfo{volume}{45},
  \bibinfo{pages}{1367--1378}.
\bibitem[{Ji et~al.(2018)Ji, Sarrazin, Leboyer, Guevara, Guevara, Poupon,
  Grigis and Houenou}]{Ji2018-hq}
\bibinfo{author}{Ji, E.}, \bibinfo{author}{Sarrazin, S.},
  \bibinfo{author}{Leboyer, M.}, \bibinfo{author}{Guevara, M.},
  \bibinfo{author}{Guevara, P.}, \bibinfo{author}{Poupon, C.},
  \bibinfo{author}{Grigis, A.}, \bibinfo{author}{Houenou, J.},
  \bibinfo{year}{2018}.
\newblock \bibinfo{title}{T240. relationship between cognitive performance and
  superficial white matter integrity in the cingulate cortex in schizophrenia:
  A {DWI} study using a novel atlas}.
\newblock \bibinfo{journal}{Biol. Psychiatry} \bibinfo{volume}{83},
  \bibinfo{pages}{S222}.
\bibitem[{Jin et~al.(2014)Jin, Shi, Zhan, Gutman, de~Zubicaray, McMahon,
  Wright, Toga and Thompson}]{Jin2014-wv}
\bibinfo{author}{Jin, Y.}, \bibinfo{author}{Shi, Y.}, \bibinfo{author}{Zhan,
  L.}, \bibinfo{author}{Gutman, B.A.}, \bibinfo{author}{de~Zubicaray, G.I.},
  \bibinfo{author}{McMahon, K.L.}, \bibinfo{author}{Wright, M.J.},
  \bibinfo{author}{Toga, A.W.}, \bibinfo{author}{Thompson, P.M.},
  \bibinfo{year}{2014}.
\newblock \bibinfo{title}{Automatic clustering of white matter fibers in brain
  diffusion {MRI} with an application to genetics}.
\newblock \bibinfo{journal}{Neuroimage} \bibinfo{volume}{100},
  \bibinfo{pages}{75--90}.
\bibitem[{Khosla et~al.(2020)Khosla, Teterwak, Wang, Sarna, Tian, Isola,
  Maschinot, Liu and Krishnan}]{Khosla2020-mx}
\bibinfo{author}{Khosla, P.}, \bibinfo{author}{Teterwak, P.},
  \bibinfo{author}{Wang, C.}, \bibinfo{author}{Sarna, A.},
  \bibinfo{author}{Tian, Y.}, \bibinfo{author}{Isola, P.},
  \bibinfo{author}{Maschinot, A.}, \bibinfo{author}{Liu, C.},
  \bibinfo{author}{Krishnan, D.}, \bibinfo{year}{2020}.
\newblock \bibinfo{title}{Supervised contrastive learning}.
\newblock \bibinfo{journal}{Advances in Neural Information Processing Systems
  (NeurIPS)} \bibinfo{volume}{33}, \bibinfo{pages}{18661--18673}.
\bibitem[{Kingma and Ba(2015)}]{Kingma2014-hv}
\bibinfo{author}{Kingma, D.P.}, \bibinfo{author}{Ba, J.}, \bibinfo{year}{2015}.
\newblock \bibinfo{title}{Adam: A method for stochastic optimization}, in:
  \bibinfo{booktitle}{International Conference on Learning Representations
  (ICLR)}.
\bibitem[{Kopuklu et~al.(2021)Kopuklu, Zheng, Xu and Rigoll}]{Kopuklu2021-vt}
\bibinfo{author}{Kopuklu, O.}, \bibinfo{author}{Zheng, J.},
  \bibinfo{author}{Xu, H.}, \bibinfo{author}{Rigoll, G.}, \bibinfo{year}{2021}.
\newblock \bibinfo{title}{Driver anomaly detection: A dataset and contrastive
  learning approach}, in: \bibinfo{booktitle}{2021 {IEEE} Winter Conference on
  Applications of Computer Vision ({WACV})}.
\bibitem[{Lin et~al.(2017)Lin, Goyal, Girshick, He and Doll{\'a}r}]{Lin2017-or}
\bibinfo{author}{Lin, T.Y.}, \bibinfo{author}{Goyal, P.},
  \bibinfo{author}{Girshick, R.}, \bibinfo{author}{He, K.},
  \bibinfo{author}{Doll{\'a}r, P.}, \bibinfo{year}{2017}.
\newblock \bibinfo{title}{Focal loss for dense object detection}, in:
  \bibinfo{booktitle}{Proceedings of the IEEE international conference on
  computer vision (ICCV)}, pp. \bibinfo{pages}{2980--2988}.
\bibitem[{Liu et~al.(2019)Liu, Feng, Chen, Wu, Hong, Yap and Shen}]{Liu2019-jj}
\bibinfo{author}{Liu, F.}, \bibinfo{author}{Feng, J.}, \bibinfo{author}{Chen,
  G.}, \bibinfo{author}{Wu, Y.}, \bibinfo{author}{Hong, Y.},
  \bibinfo{author}{Yap, P.T.}, \bibinfo{author}{Shen, D.},
  \bibinfo{year}{2019}.
\newblock \bibinfo{title}{{DeepBundle}: Fiber bundle parcellation with graph
  convolution neural networks}, in: \bibinfo{booktitle}{Graph Learning in
  Medical Imaging}, pp. \bibinfo{pages}{88--95}.
\bibitem[{Lu et~al.(2021)Lu, Li and Ye}]{Lu2021-hr}
\bibinfo{author}{Lu, Q.}, \bibinfo{author}{Li, Y.}, \bibinfo{author}{Ye, C.},
  \bibinfo{year}{2021}.
\newblock \bibinfo{title}{Volumetric white matter tract segmentation with
  nested self-supervised learning using sequential pretext tasks}.
\newblock \bibinfo{journal}{Med. Image Anal.} \bibinfo{volume}{72},
  \bibinfo{pages}{102094}.
\bibitem[{Lu et~al.(2022)Lu, Liu, Zhuo, Li, Duan, Yu, Qu, Ye and
  Liu}]{Lu2022-ym}
\bibinfo{author}{Lu, Q.}, \bibinfo{author}{Liu, W.}, \bibinfo{author}{Zhuo,
  Z.}, \bibinfo{author}{Li, Y.}, \bibinfo{author}{Duan, Y.},
  \bibinfo{author}{Yu, P.}, \bibinfo{author}{Qu, L.}, \bibinfo{author}{Ye, C.},
  \bibinfo{author}{Liu, Y.}, \bibinfo{year}{2022}.
\newblock \bibinfo{title}{A transfer learning approach to few-shot segmentation
  of novel white matter tracts}.
\newblock \bibinfo{journal}{Med. Image Anal.} \bibinfo{volume}{79},
  \bibinfo{pages}{102454}.
\bibitem[{Malcolm et~al.(2010)Malcolm, Shenton and Rathi}]{Malcolm2010-mk}
\bibinfo{author}{Malcolm, J.G.}, \bibinfo{author}{Shenton, M.E.},
  \bibinfo{author}{Rathi, Y.}, \bibinfo{year}{2010}.
\newblock \bibinfo{title}{Filtered multitensor tractography}.
\newblock \bibinfo{journal}{IEEE Trans. Med. Imaging} \bibinfo{volume}{29},
  \bibinfo{pages}{1664--1675}.
\bibitem[{Malykhin et~al.(2011)Malykhin, Vahidy, Michielse, Coupland,
  Camicioli, Seres and Carter}]{Malykhin2011-ex}
\bibinfo{author}{Malykhin, N.}, \bibinfo{author}{Vahidy, S.},
  \bibinfo{author}{Michielse, S.}, \bibinfo{author}{Coupland, N.},
  \bibinfo{author}{Camicioli, R.}, \bibinfo{author}{Seres, P.},
  \bibinfo{author}{Carter, R.}, \bibinfo{year}{2011}.
\newblock \bibinfo{title}{Structural organization of the prefrontal white
  matter pathways in the adult and aging brain measured by diffusion tensor
  imaging}.
\newblock \bibinfo{journal}{Brain Struct. Funct.} \bibinfo{volume}{216},
  \bibinfo{pages}{417--431}.
\bibitem[{Marek et~al.(2011)Marek, Jennings, Lasch, Siderowf, Tanner, Simuni,
  Coffey, Kieburtz, Flagg, Chowdhury
  et~al.}]{Parkinson_Progression_Marker_Initiative2011-oz}
\bibinfo{author}{Marek, K.}, \bibinfo{author}{Jennings, D.},
  \bibinfo{author}{Lasch, S.}, \bibinfo{author}{Siderowf, A.},
  \bibinfo{author}{Tanner, C.}, \bibinfo{author}{Simuni, T.},
  \bibinfo{author}{Coffey, C.}, \bibinfo{author}{Kieburtz, K.},
  \bibinfo{author}{Flagg, E.}, \bibinfo{author}{Chowdhury, S.}, et~al.,
  \bibinfo{year}{2011}.
\newblock \bibinfo{title}{The parkinson progression marker initiative
  ({PPMI})}.
\newblock \bibinfo{journal}{Prog. Neurobiol.} \bibinfo{volume}{95},
  \bibinfo{pages}{629--635}.
\bibitem[{Ngattai~Lam et~al.(2018)Ngattai~Lam, Belhomme, Ferrall, Patterson,
  Styner and Prieto}]{Ngattai_Lam2018-ne}
\bibinfo{author}{Ngattai~Lam, P.D.}, \bibinfo{author}{Belhomme, G.},
  \bibinfo{author}{Ferrall, J.}, \bibinfo{author}{Patterson, B.},
  \bibinfo{author}{Styner, M.}, \bibinfo{author}{Prieto, J.C.},
  \bibinfo{year}{2018}.
\newblock \bibinfo{title}{{TRAFIC}: Fiber tract classification using deep
  learning}.
\newblock \bibinfo{journal}{Proc. SPIE Int. Soc. Opt. Eng.}
  \bibinfo{volume}{10574}.
\bibitem[{Norton et~al.(2017)Norton, Essayed, Zhang, Pujol, Yarmarkovich,
  Golby, Kindlmann, Wassermann, Estepar, Rathi, Pieper, Kikinis, Johnson,
  Westin and O'Donnell}]{Norton2017-nz}
\bibinfo{author}{Norton, I.}, \bibinfo{author}{Essayed, W.I.},
  \bibinfo{author}{Zhang, F.}, \bibinfo{author}{Pujol, S.},
  \bibinfo{author}{Yarmarkovich, A.}, \bibinfo{author}{Golby, A.J.},
  \bibinfo{author}{Kindlmann, G.}, \bibinfo{author}{Wassermann, D.},
  \bibinfo{author}{Estepar, R.S.J.}, \bibinfo{author}{Rathi, Y.},
  \bibinfo{author}{Pieper, S.}, \bibinfo{author}{Kikinis, R.},
  \bibinfo{author}{Johnson, H.J.}, \bibinfo{author}{Westin, C.F.},
  \bibinfo{author}{O'Donnell, L.J.}, \bibinfo{year}{2017}.
\newblock \bibinfo{title}{{SlicerDMRI}: Open source diffusion {MRI} software
  for brain cancer research}.
\newblock \bibinfo{journal}{Cancer Res.} \bibinfo{volume}{77},
  \bibinfo{pages}{e101--e103}.
\bibitem[{O'Donnell et~al.(2017)O'Donnell, Suter, Rigolo, Kahali, Zhang,
  Norton, Albi, Olubiyi, Meola, Essayed, Unadkat, Ciris, Wells, Rathi, Westin
  and Golby}]{ODonnell2017-sq}
\bibinfo{author}{O'Donnell, L.J.}, \bibinfo{author}{Suter, Y.},
  \bibinfo{author}{Rigolo, L.}, \bibinfo{author}{Kahali, P.},
  \bibinfo{author}{Zhang, F.}, \bibinfo{author}{Norton, I.},
  \bibinfo{author}{Albi, A.}, \bibinfo{author}{Olubiyi, O.},
  \bibinfo{author}{Meola, A.}, \bibinfo{author}{Essayed, W.I.},
  \bibinfo{author}{Unadkat, P.}, \bibinfo{author}{Ciris, P.A.},
  \bibinfo{author}{Wells, W.M.}, \bibinfo{author}{Rathi, Y.},
  \bibinfo{author}{Westin, C.F.}, \bibinfo{author}{Golby, A.J.},
  \bibinfo{year}{2017}.
\newblock \bibinfo{title}{Automated white matter fiber tract identification in
  patients with brain tumors}.
\newblock \bibinfo{journal}{NeuroImage: Clinical} \bibinfo{volume}{13},
  \bibinfo{pages}{138--153}.
\bibitem[{O'Donnell et~al.(2012)O'Donnell, Wells, Golby and
  Westin}]{ODonnell2012-ol}
\bibinfo{author}{O'Donnell, L.J.}, \bibinfo{author}{Wells, 3rd, W.M.},
  \bibinfo{author}{Golby, A.J.}, \bibinfo{author}{Westin, C.F.},
  \bibinfo{year}{2012}.
\newblock \bibinfo{title}{Unbiased groupwise registration of white matter
  tractography}, in: \bibinfo{booktitle}{International Conference on Medical
  Image Computing and Computer Assisted Intervention (MICCAI)}, pp.
  \bibinfo{pages}{123--130}.
\bibitem[{O'Donnell and Westin(2007)}]{ODonnell2007-sa}
\bibinfo{author}{O'Donnell, L.J.}, \bibinfo{author}{Westin, C.F.},
  \bibinfo{year}{2007}.
\newblock \bibinfo{title}{Automatic tractography segmentation using a
  high-dimensional white matter atlas}.
\newblock \bibinfo{journal}{IEEE Trans. Med. Imaging} \bibinfo{volume}{26},
  \bibinfo{pages}{1562--1575}.
\bibitem[{Oishi et~al.(2008)Oishi, Zilles, Amunts, Faria, Jiang, Li, Akhter,
  Hua, Woods, Toga, Pike, Rosa-Neto, Evans, Zhang, Huang, Miller, van Zijl,
  Mazziotta and Mori}]{Oishi2008-ct}
\bibinfo{author}{Oishi, K.}, \bibinfo{author}{Zilles, K.},
  \bibinfo{author}{Amunts, K.}, \bibinfo{author}{Faria, A.},
  \bibinfo{author}{Jiang, H.}, \bibinfo{author}{Li, X.},
  \bibinfo{author}{Akhter, K.}, \bibinfo{author}{Hua, K.},
  \bibinfo{author}{Woods, R.}, \bibinfo{author}{Toga, A.W.},
  \bibinfo{author}{Pike, G.B.}, \bibinfo{author}{Rosa-Neto, P.},
  \bibinfo{author}{Evans, A.}, \bibinfo{author}{Zhang, J.},
  \bibinfo{author}{Huang, H.}, \bibinfo{author}{Miller, M.I.},
  \bibinfo{author}{van Zijl, P.C.M.}, \bibinfo{author}{Mazziotta, J.},
  \bibinfo{author}{Mori, S.}, \bibinfo{year}{2008}.
\newblock \bibinfo{title}{Human brain white matter atlas: identification and
  assignment of common anatomical structures in superficial white matter}.
\newblock \bibinfo{journal}{Neuroimage} \bibinfo{volume}{43},
  \bibinfo{pages}{447--457}.
\bibitem[{Ouyang et~al.(2016)Ouyang, Jeon, Mishra, Du, Wang, Peng and
  Huang}]{Ouyang2016-kj}
\bibinfo{author}{Ouyang, M.}, \bibinfo{author}{Jeon, T.},
  \bibinfo{author}{Mishra, V.}, \bibinfo{author}{Du, H.},
  \bibinfo{author}{Wang, Y.}, \bibinfo{author}{Peng, Y.},
  \bibinfo{author}{Huang, H.}, \bibinfo{year}{2016}.
\newblock \bibinfo{title}{Global and regional cortical connectivity maturation
  index ({CCMI}) of developmental human brain with quantification of
  short-range association tracts}.
\newblock \bibinfo{journal}{Proc. SPIE Int. Soc. Opt. Eng.}
  \bibinfo{volume}{9788}.
\bibitem[{Panda et~al.(2021)Panda, Korfiatis, Suman, Garg, Polley, Singh, Chari
  and Goenka}]{Panda2021-tt}
\bibinfo{author}{Panda, A.}, \bibinfo{author}{Korfiatis, P.},
  \bibinfo{author}{Suman, G.}, \bibinfo{author}{Garg, S.K.},
  \bibinfo{author}{Polley, E.C.}, \bibinfo{author}{Singh, D.P.},
  \bibinfo{author}{Chari, S.T.}, \bibinfo{author}{Goenka, A.H.},
  \bibinfo{year}{2021}.
\newblock \bibinfo{title}{Two-stage deep learning model for fully automated
  pancreas segmentation on computed tomography: Comparison with intra-reader
  and inter-reader reliability at full and reduced radiation dose on an
  external dataset}.
\newblock \bibinfo{journal}{Med. Phys.} \bibinfo{volume}{48},
  \bibinfo{pages}{2468--2481}.
\bibitem[{Poldrack et~al.(2016)Poldrack, Congdon, Triplett, Gorgolewski,
  Karlsgodt, Mumford, Sabb, Freimer, London, Cannon and
  Bilder}]{Poldrack2016-zs}
\bibinfo{author}{Poldrack, R.A.}, \bibinfo{author}{Congdon, E.},
  \bibinfo{author}{Triplett, W.}, \bibinfo{author}{Gorgolewski, K.J.},
  \bibinfo{author}{Karlsgodt, K.H.}, \bibinfo{author}{Mumford, J.A.},
  \bibinfo{author}{Sabb, F.W.}, \bibinfo{author}{Freimer, N.B.},
  \bibinfo{author}{London, E.D.}, \bibinfo{author}{Cannon, T.D.},
  \bibinfo{author}{Bilder, R.M.}, \bibinfo{year}{2016}.
\newblock \bibinfo{title}{A phenome-wide examination of neural and cognitive
  function}.
\newblock \bibinfo{journal}{Sci Data} \bibinfo{volume}{3},
  \bibinfo{pages}{160110}.
\bibitem[{Qi et~al.(2017)Qi, Yi, Su and Guibas}]{Qi2017-dw}
\bibinfo{author}{Qi, C.R.}, \bibinfo{author}{Yi, L.}, \bibinfo{author}{Su, H.},
  \bibinfo{author}{Guibas, L.J.}, \bibinfo{year}{2017}.
\newblock \bibinfo{title}{Pointnet++: Deep hierarchical feature learning on
  point sets in a metric space}.
\newblock \bibinfo{journal}{Advances in neural information processing systems
  (NeurIPS)} \bibinfo{volume}{30}.
\bibitem[{Ramos-Llord{\'e}n et~al.(2020)Ramos-Llord{\'e}n, Ning, Liao,
  Mukhometzianov, Michailovich, Setsompop and Rathi}]{Ramos-Llorden2020-vt}
\bibinfo{author}{Ramos-Llord{\'e}n, G.}, \bibinfo{author}{Ning, L.},
  \bibinfo{author}{Liao, C.}, \bibinfo{author}{Mukhometzianov, R.},
  \bibinfo{author}{Michailovich, O.}, \bibinfo{author}{Setsompop, K.},
  \bibinfo{author}{Rathi, Y.}, \bibinfo{year}{2020}.
\newblock \bibinfo{title}{High-fidelity, accelerated whole-brain submillimeter
  in vivo diffusion {MRI} using gslider-spherical ridgelets ({gSlider-SR})}.
\newblock \bibinfo{journal}{Magn. Reson. Med.} \bibinfo{volume}{84},
  \bibinfo{pages}{1781--1795}.
\bibitem[{Reddy and Rathi(2016)}]{Reddy2016-ko}
\bibinfo{author}{Reddy, C.P.}, \bibinfo{author}{Rathi, Y.},
  \bibinfo{year}{2016}.
\newblock \bibinfo{title}{Joint {Multi-Fiber} {NODDI} parameter estimation and
  tractography using the unscented information filter}.
\newblock \bibinfo{journal}{Front. Neurosci.} \bibinfo{volume}{10},
  \bibinfo{pages}{166}.
\bibitem[{Reginold et~al.(2016)Reginold, Luedke, Itorralba, Fernandez-Ruiz,
  Islam and Garcia}]{Reginold2016-zq}
\bibinfo{author}{Reginold, W.}, \bibinfo{author}{Luedke, A.C.},
  \bibinfo{author}{Itorralba, J.}, \bibinfo{author}{Fernandez-Ruiz, J.},
  \bibinfo{author}{Islam, O.}, \bibinfo{author}{Garcia, A.},
  \bibinfo{year}{2016}.
\newblock \bibinfo{title}{Altered superficial white matter on tractography
  {MRI} in alzheimer's disease}.
\newblock \bibinfo{journal}{Dement. Geriatr. Cogn. Dis. Extra}
  \bibinfo{volume}{6}, \bibinfo{pages}{233--241}.
\bibitem[{Reveley et~al.(2015)Reveley, Seth, Pierpaoli, Silva, Yu, Saunders,
  Leopold and Ye}]{Reveley2015-ki}
\bibinfo{author}{Reveley, C.}, \bibinfo{author}{Seth, A.K.},
  \bibinfo{author}{Pierpaoli, C.}, \bibinfo{author}{Silva, A.C.},
  \bibinfo{author}{Yu, D.}, \bibinfo{author}{Saunders, R.C.},
  \bibinfo{author}{Leopold, D.A.}, \bibinfo{author}{Ye, F.Q.},
  \bibinfo{year}{2015}.
\newblock \bibinfo{title}{Superficial white matter fiber systems impede
  detection of long-range cortical connections in diffusion {MR} tractography}.
\newblock \bibinfo{journal}{Proc. Natl. Acad. Sci. U. S. A.}
  \bibinfo{volume}{112}, \bibinfo{pages}{E2820--8}.
\bibitem[{Roberts et~al.(2017)Roberts, Perry, Roberts, Mitchell and
  Breakspear}]{Roberts2017-hh}
\bibinfo{author}{Roberts, J.A.}, \bibinfo{author}{Perry, A.},
  \bibinfo{author}{Roberts, G.}, \bibinfo{author}{Mitchell, P.B.},
  \bibinfo{author}{Breakspear, M.}, \bibinfo{year}{2017}.
\newblock \bibinfo{title}{Consistency-based thresholding of the human
  connectome}.
\newblock \bibinfo{journal}{Neuroimage} \bibinfo{volume}{145},
  \bibinfo{pages}{118--129}.
\bibitem[{Rom{\'a}n et~al.(2017)Rom{\'a}n, Guevara, Valenzuela, Figueroa,
  Houenou, Duclap, Poupon, Mangin and Guevara}]{Roman2017-uy}
\bibinfo{author}{Rom{\'a}n, C.}, \bibinfo{author}{Guevara, M.},
  \bibinfo{author}{Valenzuela, R.}, \bibinfo{author}{Figueroa, M.},
  \bibinfo{author}{Houenou, J.}, \bibinfo{author}{Duclap, D.},
  \bibinfo{author}{Poupon, C.}, \bibinfo{author}{Mangin, J.F.},
  \bibinfo{author}{Guevara, P.}, \bibinfo{year}{2017}.
\newblock \bibinfo{title}{Clustering of {Whole-Brain} white matter short
  association bundles using {HARDI} data}.
\newblock \bibinfo{journal}{Front. Neuroinform.} \bibinfo{volume}{11},
  \bibinfo{pages}{73}.
\bibitem[{Rom{\'a}n et~al.(2022)Rom{\'a}n, Hern{\'a}ndez, Figueroa, Houenou,
  Poupon, Mangin and Guevara}]{Roman2022-pr}
\bibinfo{author}{Rom{\'a}n, C.}, \bibinfo{author}{Hern{\'a}ndez, C.},
  \bibinfo{author}{Figueroa, M.}, \bibinfo{author}{Houenou, J.},
  \bibinfo{author}{Poupon, C.}, \bibinfo{author}{Mangin, J.F.},
  \bibinfo{author}{Guevara, P.}, \bibinfo{year}{2022}.
\newblock \bibinfo{title}{Superficial white matter bundle atlas based on
  hierarchical fiber clustering over probabilistic tractography data}.
\newblock \bibinfo{journal}{Neuroimage} \bibinfo{volume}{262},
  \bibinfo{pages}{119550}.
\bibitem[{Rom{\'a}n et~al.(2021)Rom{\'a}n, L{\'o}pez-L{\'o}pez, Houenou,
  Poupon, Mangin, Hern{\'a}ndez and Guevara}]{Roman2021-rd}
\bibinfo{author}{Rom{\'a}n, C.}, \bibinfo{author}{L{\'o}pez-L{\'o}pez, N.},
  \bibinfo{author}{Houenou, J.}, \bibinfo{author}{Poupon, C.},
  \bibinfo{author}{Mangin, J.F.}, \bibinfo{author}{Hern{\'a}ndez, C.},
  \bibinfo{author}{Guevara, P.}, \bibinfo{year}{2021}.
\newblock \bibinfo{title}{Study of {Precentral-Postcentral} connections on hcp
  data using probabilistic tractography and fiber clustering}, in:
  \bibinfo{booktitle}{2021 {IEEE} 18th International Symposium on Biomedical
  Imaging ({ISBI})}, pp. \bibinfo{pages}{55--59}.
\bibitem[{Schiffer et~al.(2021)Schiffer, Amunts, Harmeling and
  Dickscheid}]{Schiffer2021-vl}
\bibinfo{author}{Schiffer, C.}, \bibinfo{author}{Amunts, K.},
  \bibinfo{author}{Harmeling, S.}, \bibinfo{author}{Dickscheid, T.},
  \bibinfo{year}{2021}.
\newblock \bibinfo{title}{Contrastive representation learning for whole brain
  cytoarchitectonic mapping in histological human brain sections}, in:
  \bibinfo{booktitle}{2021 {IEEE} 18th International Symposium on Biomedical
  Imaging ({ISBI})}, pp. \bibinfo{pages}{603--606}.
\bibitem[{Schilling et~al.(2018)Schilling, Gao, Janve, Stepniewska, Landman and
  Anderson}]{Schilling2018-el}
\bibinfo{author}{Schilling, K.}, \bibinfo{author}{Gao, Y.},
  \bibinfo{author}{Janve, V.}, \bibinfo{author}{Stepniewska, I.},
  \bibinfo{author}{Landman, B.A.}, \bibinfo{author}{Anderson, A.W.},
  \bibinfo{year}{2018}.
\newblock \bibinfo{title}{Confirmation of a gyral bias in diffusion {MRI} fiber
  tractography}.
\newblock \bibinfo{journal}{Hum. Brain Mapp.} \bibinfo{volume}{39},
  \bibinfo{pages}{1449--1466}.
\bibitem[{Schilling et~al.(2022)Schilling, Archer, Yeh, Rheault, Cai, Shafer,
  Resnick, Hohman, Jefferson, Anderson, Kang and Landman}]{Schilling2022-yz}
\bibinfo{author}{Schilling, K.G.}, \bibinfo{author}{Archer, D.},
  \bibinfo{author}{Yeh, F.C.}, \bibinfo{author}{Rheault, F.},
  \bibinfo{author}{Cai, L.Y.}, \bibinfo{author}{Shafer, A.},
  \bibinfo{author}{Resnick, S.M.}, \bibinfo{author}{Hohman, T.},
  \bibinfo{author}{Jefferson, A.}, \bibinfo{author}{Anderson, A.W.},
  \bibinfo{author}{Kang, H.}, \bibinfo{author}{Landman, B.A.},
  \bibinfo{year}{2022}.
\newblock \bibinfo{title}{Short superficial white matter and aging: a
  longitudinal multi-site study of 1,293 subjects and 2,711 sessions}.
\newblock \bibinfo{journal}{bioRxiv: 10.1101/2022.06.06.494720} .
\bibitem[{Song et~al.(2014)Song, Chang, Petty, Guidon and Chen}]{Song2014-un}
\bibinfo{author}{Song, A.W.}, \bibinfo{author}{Chang, H.C.},
  \bibinfo{author}{Petty, C.}, \bibinfo{author}{Guidon, A.},
  \bibinfo{author}{Chen, N.K.}, \bibinfo{year}{2014}.
\newblock \bibinfo{title}{Improved delineation of short cortical association
  fibers and gray/white matter boundary using whole-brain three-dimensional
  diffusion tensor imaging at submillimeter spatial resolution}.
\newblock \bibinfo{journal}{Brain Connect.} \bibinfo{volume}{4},
  \bibinfo{pages}{636--640}.
\bibitem[{St-Onge et~al.(2018)St-Onge, Daducci, Girard and
  Descoteaux}]{St-Onge2018-rw}
\bibinfo{author}{St-Onge, E.}, \bibinfo{author}{Daducci, A.},
  \bibinfo{author}{Girard, G.}, \bibinfo{author}{Descoteaux, M.},
  \bibinfo{year}{2018}.
\newblock \bibinfo{title}{Surface-enhanced tractography ({SET})}.
\newblock \bibinfo{journal}{Neuroimage} \bibinfo{volume}{169},
  \bibinfo{pages}{524--539}.
\bibitem[{Taha and Hanbury(2015)}]{Taha2015-ij}
\bibinfo{author}{Taha, A.A.}, \bibinfo{author}{Hanbury, A.},
  \bibinfo{year}{2015}.
\newblock \bibinfo{title}{Metrics for evaluating {3D} medical image
  segmentation: analysis, selection, and tool}.
\newblock \bibinfo{journal}{BMC Med. Imaging} \bibinfo{volume}{15},
  \bibinfo{pages}{29}.
\bibitem[{Van~Essen(2013)}]{Van_Essen2013-ka}
\bibinfo{author}{Van~Essen, D.C.}, \bibinfo{year}{2013}.
\newblock \bibinfo{title}{Cartography and connectomes}.
\newblock \bibinfo{journal}{Neuron} \bibinfo{volume}{80},
  \bibinfo{pages}{775--790}.
\bibitem[{Van~Essen et~al.(2014)Van~Essen, Jbabdi, Sotiropoulos, Chen,
  Dikranian, Coalson, Harwell, Behrens and Glasser}]{Van_Essen2014-mc}
\bibinfo{author}{Van~Essen, D.C.}, \bibinfo{author}{Jbabdi, S.},
  \bibinfo{author}{Sotiropoulos, S.N.}, \bibinfo{author}{Chen, C.},
  \bibinfo{author}{Dikranian, K.}, \bibinfo{author}{Coalson, T.},
  \bibinfo{author}{Harwell, J.}, \bibinfo{author}{Behrens, T.E.J.},
  \bibinfo{author}{Glasser, M.F.}, \bibinfo{year}{2014}.
\newblock \bibinfo{title}{Chapter 16 - mapping connections in humans and
  {Non-Human} primates: Aspirations and challenges for diffusion imaging}, in:
  \bibinfo{booktitle}{Diffusion {MRI} (Second Edition)}, pp.
  \bibinfo{pages}{337--358}.
\bibitem[{Van~Essen et~al.(2013)Van~Essen, Smith, Barch, Behrens, Yacoub,
  Ugurbil and {WU-Minn HCP Consortium}}]{Van_Essen2013-ne}
\bibinfo{author}{Van~Essen, D.C.}, \bibinfo{author}{Smith, S.M.},
  \bibinfo{author}{Barch, D.M.}, \bibinfo{author}{Behrens, T.E.J.},
  \bibinfo{author}{Yacoub, E.}, \bibinfo{author}{Ugurbil, K.},
  \bibinfo{author}{{WU-Minn HCP Consortium}}, \bibinfo{year}{2013}.
\newblock \bibinfo{title}{The {WU-Minn} human connectome project: an overview}.
\newblock \bibinfo{journal}{Neuroimage} \bibinfo{volume}{80},
  \bibinfo{pages}{62--79}.
\bibitem[{Volkow et~al.(2018)Volkow, Koob, Croyle, Bianchi, Gordon, Koroshetz,
  P{\'e}rez-Stable, Riley, Bloch, Conway, Deeds, Dowling, Grant, Howlett,
  Matochik, Morgan, Murray, Noronha, Spong, Wargo, Warren and
  Weiss}]{Volkow2018-og}
\bibinfo{author}{Volkow, N.D.}, \bibinfo{author}{Koob, G.F.},
  \bibinfo{author}{Croyle, R.T.}, \bibinfo{author}{Bianchi, D.W.},
  \bibinfo{author}{Gordon, J.A.}, \bibinfo{author}{Koroshetz, W.J.},
  \bibinfo{author}{P{\'e}rez-Stable, E.J.}, \bibinfo{author}{Riley, W.T.},
  \bibinfo{author}{Bloch, M.H.}, \bibinfo{author}{Conway, K.},
  \bibinfo{author}{Deeds, B.G.}, \bibinfo{author}{Dowling, G.J.},
  \bibinfo{author}{Grant, S.}, \bibinfo{author}{Howlett, K.D.},
  \bibinfo{author}{Matochik, J.A.}, \bibinfo{author}{Morgan, G.D.},
  \bibinfo{author}{Murray, M.M.}, \bibinfo{author}{Noronha, A.},
  \bibinfo{author}{Spong, C.Y.}, \bibinfo{author}{Wargo, E.M.},
  \bibinfo{author}{Warren, K.R.}, \bibinfo{author}{Weiss, S.R.B.},
  \bibinfo{year}{2018}.
\newblock \bibinfo{title}{The conception of the {ABCD} study: From substance
  use to a broad {NIH} collaboration}.
\newblock \bibinfo{journal}{Dev. Cogn. Neurosci.} \bibinfo{volume}{32},
  \bibinfo{pages}{4--7}.
\bibitem[{Wang et~al.(2019)Wang, Sun, Liu, Sarma, Bronstein and
  Solomon}]{Wang2018-xa}
\bibinfo{author}{Wang, Y.}, \bibinfo{author}{Sun, Y.}, \bibinfo{author}{Liu,
  Z.}, \bibinfo{author}{Sarma, S.E.}, \bibinfo{author}{Bronstein, M.M.},
  \bibinfo{author}{Solomon, J.M.}, \bibinfo{year}{2019}.
\newblock \bibinfo{title}{Dynamic graph cnn for learning on point clouds}.
\newblock \bibinfo{journal}{ACM Trans. on Graphics} \bibinfo{volume}{38},
  \bibinfo{pages}{1--12}.
\bibitem[{Wasserthal et~al.(2018)Wasserthal, Neher and
  Maier-Hein}]{Wasserthal2018-if}
\bibinfo{author}{Wasserthal, J.}, \bibinfo{author}{Neher, P.},
  \bibinfo{author}{Maier-Hein, K.H.}, \bibinfo{year}{2018}.
\newblock \bibinfo{title}{{TractSeg} - fast and accurate white matter tract
  segmentation}.
\newblock \bibinfo{journal}{Neuroimage} \bibinfo{volume}{183},
  \bibinfo{pages}{239--253}.
\bibitem[{Wasserthal et~al.(2019)Wasserthal, Neher, Hirjak and
  Maier-Hein}]{Wasserthal2019-ha}
\bibinfo{author}{Wasserthal, J.}, \bibinfo{author}{Neher, P.F.},
  \bibinfo{author}{Hirjak, D.}, \bibinfo{author}{Maier-Hein, K.H.},
  \bibinfo{year}{2019}.
\newblock \bibinfo{title}{Combined tract segmentation and orientation mapping
  for bundle-specific tractography}.
\newblock \bibinfo{journal}{Med. Image Anal.} \bibinfo{volume}{58},
  \bibinfo{pages}{101559}.
\bibitem[{Wen et~al.(2016)Wen, Zhang, Li and Qiao}]{Wen2016-zn}
\bibinfo{author}{Wen, Y.}, \bibinfo{author}{Zhang, K.}, \bibinfo{author}{Li,
  Z.}, \bibinfo{author}{Qiao, Y.}, \bibinfo{year}{2016}.
\newblock \bibinfo{title}{A discriminative feature learning approach for deep
  face recognition}, in: \bibinfo{booktitle}{European Conference on Computer
  Vision (ECCV)}, pp. \bibinfo{pages}{499--515}.
\bibitem[{Wu et~al.(2014)Wu, Lu, Lowes, Yang, Passarotti, Zhou and
  Pavuluri}]{Wu2014-ai}
\bibinfo{author}{Wu, M.}, \bibinfo{author}{Lu, L.H.}, \bibinfo{author}{Lowes,
  A.}, \bibinfo{author}{Yang, S.}, \bibinfo{author}{Passarotti, A.M.},
  \bibinfo{author}{Zhou, X.J.}, \bibinfo{author}{Pavuluri, M.N.},
  \bibinfo{year}{2014}.
\newblock \bibinfo{title}{Development of superficial white matter and its
  structural interplay with cortical gray matter in children and adolescents}.
\newblock \bibinfo{journal}{Hum. Brain Mapp.} \bibinfo{volume}{35},
  \bibinfo{pages}{2806--2816}.
\bibitem[{Wu et~al.(2020a)Wu, Phang, Park, Shen, Huang, Zorin, Jastrzebski,
  Fevry, Katsnelson, Kim, Wolfson, Parikh, Gaddam, Lin, Ho, Weinstein, Reig,
  Gao, Toth, Pysarenko, Lewin, Lee, Airola, Mema, Chung, Hwang, Samreen, Kim,
  Heacock, Moy, Cho and Geras}]{Wu2020-va}
\bibinfo{author}{Wu, N.}, \bibinfo{author}{Phang, J.}, \bibinfo{author}{Park,
  J.}, \bibinfo{author}{Shen, Y.}, \bibinfo{author}{Huang, Z.},
  \bibinfo{author}{Zorin, M.}, \bibinfo{author}{Jastrzebski, S.},
  \bibinfo{author}{Fevry, T.}, \bibinfo{author}{Katsnelson, J.},
  \bibinfo{author}{Kim, E.}, \bibinfo{author}{Wolfson, S.},
  \bibinfo{author}{Parikh, U.}, \bibinfo{author}{Gaddam, S.},
  \bibinfo{author}{Lin, L.L.Y.}, \bibinfo{author}{Ho, K.},
  \bibinfo{author}{Weinstein, J.D.}, \bibinfo{author}{Reig, B.},
  \bibinfo{author}{Gao, Y.}, \bibinfo{author}{Toth, H.},
  \bibinfo{author}{Pysarenko, K.}, \bibinfo{author}{Lewin, A.},
  \bibinfo{author}{Lee, J.}, \bibinfo{author}{Airola, K.},
  \bibinfo{author}{Mema, E.}, \bibinfo{author}{Chung, S.},
  \bibinfo{author}{Hwang, E.}, \bibinfo{author}{Samreen, N.},
  \bibinfo{author}{Kim, S.G.}, \bibinfo{author}{Heacock, L.},
  \bibinfo{author}{Moy, L.}, \bibinfo{author}{Cho, K.}, \bibinfo{author}{Geras,
  K.J.}, \bibinfo{year}{2020}a.
\newblock \bibinfo{title}{Deep neural networks improve radiologists'
  performance in breast cancer screening}.
\newblock \bibinfo{journal}{IEEE Trans. Med. Imaging} \bibinfo{volume}{39},
  \bibinfo{pages}{1184--1194}.
\bibitem[{Wu et~al.(2020b)Wu, Hong, Feng, Shen and Yap}]{Wu2020-hn}
\bibinfo{author}{Wu, Y.}, \bibinfo{author}{Hong, Y.}, \bibinfo{author}{Feng,
  Y.}, \bibinfo{author}{Shen, D.}, \bibinfo{author}{Yap, P.T.},
  \bibinfo{year}{2020}b.
\newblock \bibinfo{title}{Mitigating gyral bias in cortical tractography via
  asymmetric fiber orientation distributions}.
\newblock \bibinfo{journal}{Med. Image Anal.} \bibinfo{volume}{59},
  \bibinfo{pages}{101543}.
\bibitem[{Xu et~al.(2019)Xu, Dong, Lee, OrHara, Asano and Jeong}]{Xu2019-aa}
\bibinfo{author}{Xu, H.}, \bibinfo{author}{Dong, M.}, \bibinfo{author}{Lee,
  M.H.}, \bibinfo{author}{OrHara, N.}, \bibinfo{author}{Asano, E.},
  \bibinfo{author}{Jeong, J.W.}, \bibinfo{year}{2019}.
\newblock \bibinfo{title}{Objective detection of eloquent axonal pathways to
  minimize postoperative deficits in pediatric epilepsy surgery using diffusion
  tractography and convolutional neural networks}.
\newblock \bibinfo{journal}{IEEE Trans. Med. Imaging} .
\bibitem[{Xu et~al.(2018)Xu, Dong, Nakai, Asano and Jeong}]{Xu2018-rk}
\bibinfo{author}{Xu, H.}, \bibinfo{author}{Dong, M.}, \bibinfo{author}{Nakai,
  Y.}, \bibinfo{author}{Asano, E.}, \bibinfo{author}{Jeong, J.W.},
  \bibinfo{year}{2018}.
\newblock \bibinfo{title}{Automatic detection of eloquent axonal pathways in
  diffusion tractography using intracanial electrical stimulation mapping and
  convolutional neural networks}, in: \bibinfo{booktitle}{2018 {IEEE} 15th
  International Symposium on Biomedical Imaging ({ISBI})}, pp.
  \bibinfo{pages}{1034--1037}.
\bibitem[{Xu et~al.(2015)Xu, Ba, Kiros, Cho, Courville, Salakhutdinov, Zemel
  and Bengio}]{Xu2015-ns}
\bibinfo{author}{Xu, K.}, \bibinfo{author}{Ba, J.}, \bibinfo{author}{Kiros,
  R.}, \bibinfo{author}{Cho, K.}, \bibinfo{author}{Courville, A.},
  \bibinfo{author}{Salakhutdinov, R.}, \bibinfo{author}{Zemel, R.},
  \bibinfo{author}{Bengio, Y.}, \bibinfo{year}{2015}.
\newblock \bibinfo{title}{Show, attend and tell: Neural image caption
  generation with visual attention}, in: \bibinfo{booktitle}{International
  conference on machine learning (ICML)}, pp. \bibinfo{pages}{2048--2057}.
\bibitem[{Xue et~al.(2022)Xue, Zhang, Zhang, Chen, Song, Makris, Rathi, Cai and
  O'Donnell}]{Xue2022-yz}
\bibinfo{author}{Xue, T.}, \bibinfo{author}{Zhang, F.}, \bibinfo{author}{Zhang,
  C.}, \bibinfo{author}{Chen, Y.}, \bibinfo{author}{Song, Y.},
  \bibinfo{author}{Makris, N.}, \bibinfo{author}{Rathi, Y.},
  \bibinfo{author}{Cai, W.}, \bibinfo{author}{O'Donnell, L.J.},
  \bibinfo{year}{2022}.
\newblock \bibinfo{title}{Supwma: Consistent and efficient tractography
  parcellation of superficial white matter with deep learning}, in:
  \bibinfo{booktitle}{2022 {IEEE} 19th International Symposium on Biomedical
  Imaging ({ISBI})}, pp. \bibinfo{pages}{1--5}.
\bibitem[{Yendiki et~al.(2011)Yendiki, Panneck, Srinivasan, Stevens,
  Z{\"o}llei, Augustinack, Wang, Salat, Ehrlich, Behrens, Jbabdi, Gollub and
  Fischl}]{Yendiki2011-ea}
\bibinfo{author}{Yendiki, A.}, \bibinfo{author}{Panneck, P.},
  \bibinfo{author}{Srinivasan, P.}, \bibinfo{author}{Stevens, A.},
  \bibinfo{author}{Z{\"o}llei, L.}, \bibinfo{author}{Augustinack, J.},
  \bibinfo{author}{Wang, R.}, \bibinfo{author}{Salat, D.},
  \bibinfo{author}{Ehrlich, S.}, \bibinfo{author}{Behrens, T.},
  \bibinfo{author}{Jbabdi, S.}, \bibinfo{author}{Gollub, R.},
  \bibinfo{author}{Fischl, B.}, \bibinfo{year}{2011}.
\newblock \bibinfo{title}{Automated probabilistic reconstruction of
  white-matter pathways in health and disease using an atlas of the underlying
  anatomy}.
\newblock \bibinfo{journal}{Front. Neuroinform.} \bibinfo{volume}{5},
  \bibinfo{pages}{23}.
\bibitem[{Zhang et~al.(2020a)Zhang, Cetin~Karayumak, Hoffmann, Rathi, Golby and
  O'Donnell}]{Zhang2020-vm}
\bibinfo{author}{Zhang, F.}, \bibinfo{author}{Cetin~Karayumak, S.},
  \bibinfo{author}{Hoffmann, N.}, \bibinfo{author}{Rathi, Y.},
  \bibinfo{author}{Golby, A.J.}, \bibinfo{author}{O'Donnell, L.J.},
  \bibinfo{year}{2020}a.
\newblock \bibinfo{title}{Deep white matter analysis ({DeepWMA)}: Fast and
  consistent tractography segmentation}.
\newblock \bibinfo{journal}{Med. Image Anal.} \bibinfo{volume}{65},
  \bibinfo{pages}{101761}.
\bibitem[{Zhang et~al.(2022a)Zhang, Daducci, He, Schiavi, Seguin, Smith, Yeh,
  Zhao and O'Donnell}]{Zhang2022-sj}
\bibinfo{author}{Zhang, F.}, \bibinfo{author}{Daducci, A.},
  \bibinfo{author}{He, Y.}, \bibinfo{author}{Schiavi, S.},
  \bibinfo{author}{Seguin, C.}, \bibinfo{author}{Smith, R.E.},
  \bibinfo{author}{Yeh, C.H.}, \bibinfo{author}{Zhao, T.},
  \bibinfo{author}{O'Donnell, L.J.}, \bibinfo{year}{2022}a.
\newblock \bibinfo{title}{Quantitative mapping of the brain's structural
  connectivity using diffusion {MRI} tractography: A review}.
\newblock \bibinfo{journal}{Neuroimage} \bibinfo{volume}{249},
  \bibinfo{pages}{118870}.
\bibitem[{Zhang et~al.(2020b)Zhang, Noh, Juvekar, Frisken, Rigolo, Norton,
  Kapur, Pujol, Wells, Yarmarkovich, Kindlmann, Wassermann, San Jose~Estepar,
  Rathi, Kikinis, Johnson, Westin, Pieper, Golby and O'Donnell}]{Zhang2020-cv}
\bibinfo{author}{Zhang, F.}, \bibinfo{author}{Noh, T.},
  \bibinfo{author}{Juvekar, P.}, \bibinfo{author}{Frisken, S.F.},
  \bibinfo{author}{Rigolo, L.}, \bibinfo{author}{Norton, I.},
  \bibinfo{author}{Kapur, T.}, \bibinfo{author}{Pujol, S.},
  \bibinfo{author}{Wells, 3rd, W.}, \bibinfo{author}{Yarmarkovich, A.},
  \bibinfo{author}{Kindlmann, G.}, \bibinfo{author}{Wassermann, D.},
  \bibinfo{author}{San Jose~Estepar, R.}, \bibinfo{author}{Rathi, Y.},
  \bibinfo{author}{Kikinis, R.}, \bibinfo{author}{Johnson, H.J.},
  \bibinfo{author}{Westin, C.F.}, \bibinfo{author}{Pieper, S.},
  \bibinfo{author}{Golby, A.J.}, \bibinfo{author}{O'Donnell, L.J.},
  \bibinfo{year}{2020}b.
\newblock \bibinfo{title}{{SlicerDMRI}: Diffusion {MRI} and tractography
  research software for brain cancer surgery planning and visualization}.
\newblock \bibinfo{journal}{JCO Clin Cancer Inform} \bibinfo{volume}{4},
  \bibinfo{pages}{299--309}.
\bibitem[{Zhang et~al.(2017a)Zhang, Norton, Cai, Song, Wells and
  O'Donnell}]{Zhang2017-ui}
\bibinfo{author}{Zhang, F.}, \bibinfo{author}{Norton, I.},
  \bibinfo{author}{Cai, W.}, \bibinfo{author}{Song, Y.},
  \bibinfo{author}{Wells, W.M.}, \bibinfo{author}{O'Donnell, L.J.},
  \bibinfo{year}{2017}a.
\newblock \bibinfo{title}{Comparison between two white matter segmentation
  strategies: An investigation into white matter segmentation consistency}, in:
  \bibinfo{booktitle}{2017 {IEEE} 14th International Symposium on Biomedical
  Imaging ({ISBI})}, pp. \bibinfo{pages}{796--799}.
\bibitem[{Zhang et~al.(2018a)Zhang, Savadjiev, Cai, Song, Rathi, Tun{\c c},
  Parker, Kapur, Schultz, Makris, Verma and O'Donnell}]{Zhang2018-gd}
\bibinfo{author}{Zhang, F.}, \bibinfo{author}{Savadjiev, P.},
  \bibinfo{author}{Cai, W.}, \bibinfo{author}{Song, Y.},
  \bibinfo{author}{Rathi, Y.}, \bibinfo{author}{Tun{\c c}, B.},
  \bibinfo{author}{Parker, D.}, \bibinfo{author}{Kapur, T.},
  \bibinfo{author}{Schultz, R.T.}, \bibinfo{author}{Makris, N.},
  \bibinfo{author}{Verma, R.}, \bibinfo{author}{O'Donnell, L.J.},
  \bibinfo{year}{2018}a.
\newblock \bibinfo{title}{Whole brain white matter connectivity analysis using
  machine learning: An application to autism}.
\newblock \bibinfo{journal}{Neuroimage} \bibinfo{volume}{172},
  \bibinfo{pages}{826--837}.
\bibitem[{Zhang et~al.(2022b)Zhang, Wells and O'Donnell}]{Zhang2022-vc}
\bibinfo{author}{Zhang, F.}, \bibinfo{author}{Wells, W.M.},
  \bibinfo{author}{O'Donnell, L.J.}, \bibinfo{year}{2022}b.
\newblock \bibinfo{title}{Deep diffusion {MRI} registration ({DDMReg)}: A deep
  learning method for diffusion {MRI} registration}.
\newblock \bibinfo{journal}{IEEE Trans. Med. Imaging} \bibinfo{volume}{41},
  \bibinfo{pages}{1454--1467}.
\bibitem[{Zhang et~al.(2019a)Zhang, Wu, Norton, Rathi, Golby and
  O'Donnell}]{Zhang2019-bl}
\bibinfo{author}{Zhang, F.}, \bibinfo{author}{Wu, Y.}, \bibinfo{author}{Norton,
  I.}, \bibinfo{author}{Rathi, Y.}, \bibinfo{author}{Golby, A.J.},
  \bibinfo{author}{O'Donnell, L.J.}, \bibinfo{year}{2019}a.
\newblock \bibinfo{title}{Test-retest reproducibility of white matter
  parcellation using diffusion {MRI} tractography fiber clustering}.
\newblock \bibinfo{journal}{Hum. Brain Mapp.} \bibinfo{volume}{40},
  \bibinfo{pages}{3041--3057}.
\bibitem[{Zhang et~al.(2018b)Zhang, Wu, Norton, Rigolo, Rathi, Makris and
  O'Donnell}]{Zhang2018-jx}
\bibinfo{author}{Zhang, F.}, \bibinfo{author}{Wu, Y.}, \bibinfo{author}{Norton,
  I.}, \bibinfo{author}{Rigolo, L.}, \bibinfo{author}{Rathi, Y.},
  \bibinfo{author}{Makris, N.}, \bibinfo{author}{O'Donnell, L.J.},
  \bibinfo{year}{2018}b.
\newblock \bibinfo{title}{An anatomically curated fiber clustering white matter
  atlas for consistent white matter tract parcellation across the lifespan}.
\newblock \bibinfo{journal}{Neuroimage} \bibinfo{volume}{179},
  \bibinfo{pages}{429--447}.
\bibitem[{Zhang et~al.(2017b)Zhang, Liu and Shen}]{Zhang2017-ei}
\bibinfo{author}{Zhang, J.}, \bibinfo{author}{Liu, M.}, \bibinfo{author}{Shen,
  D.}, \bibinfo{year}{2017}b.
\newblock \bibinfo{title}{Detecting anatomical landmarks from limited medical
  imaging data using {Two-Stage} {Task-Oriented} deep neural networks}.
\newblock \bibinfo{journal}{IEEE Trans. Image Process.} \bibinfo{volume}{26},
  \bibinfo{pages}{4753--4764}.
\bibitem[{Zhang et~al.(2019b)Zhang, Shi, Wei, Han, Zhu and Liu}]{Zhang2019-oq}
\bibinfo{author}{Zhang, Y.}, \bibinfo{author}{Shi, J.}, \bibinfo{author}{Wei,
  H.}, \bibinfo{author}{Han, V.}, \bibinfo{author}{Zhu, W.Z.},
  \bibinfo{author}{Liu, C.}, \bibinfo{year}{2019}b.
\newblock \bibinfo{title}{Neonate and infant brain development from birth to 2
  years assessed using {MRI-based} quantitative susceptibility mapping}.
\newblock \bibinfo{journal}{Neuroimage} \bibinfo{volume}{185},
  \bibinfo{pages}{349--360}.
\bibitem[{Zhong et~al.(2021)Zhong, Li, Wu, Ren, Kim, Kim, Buch, Neumark, Bizzo,
  Tak, Park, Lee, Kang, Park, Kim, Chung, Guo, Dayan, Kalra and
  Li}]{Zhong2021-od}
\bibinfo{author}{Zhong, A.}, \bibinfo{author}{Li, X.}, \bibinfo{author}{Wu,
  D.}, \bibinfo{author}{Ren, H.}, \bibinfo{author}{Kim, K.},
  \bibinfo{author}{Kim, Y.}, \bibinfo{author}{Buch, V.},
  \bibinfo{author}{Neumark, N.}, \bibinfo{author}{Bizzo, B.},
  \bibinfo{author}{Tak, W.Y.}, \bibinfo{author}{Park, S.Y.},
  \bibinfo{author}{Lee, Y.R.}, \bibinfo{author}{Kang, M.K.},
  \bibinfo{author}{Park, J.G.}, \bibinfo{author}{Kim, B.S.},
  \bibinfo{author}{Chung, W.J.}, \bibinfo{author}{Guo, N.},
  \bibinfo{author}{Dayan, I.}, \bibinfo{author}{Kalra, M.K.},
  \bibinfo{author}{Li, Q.}, \bibinfo{year}{2021}.
\newblock \bibinfo{title}{Deep metric learning-based image retrieval system for
  chest radiograph and its clinical applications in {COVID-19}}.
\newblock \bibinfo{journal}{Med. Image Anal.} \bibinfo{volume}{70},
  \bibinfo{pages}{101993}.

\end{thebibliography}



\end{document}